\documentclass{article}

\usepackage[main,final,nonatbib]{neurips_2026}
\usepackage[T1]{fontenc}  %
\usepackage[utf8]{inputenc}  %
\usepackage[english]{babel}  %
\usepackage{mathtools}

\usepackage{algpseudocode}
\usepackage{array}
\usepackage{booktabs}
\usepackage[caption=false,font=scriptsize]{subfig}
\usepackage{textcomp}
\usepackage{url}
\usepackage{verbatim}
\usepackage{graphicx}
\usepackage{wrapfig}
\usepackage[colorlinks=true,citecolor=blue]{hyperref}

\usepackage{import}
\usepackage{comment}
\usepackage{xcolor}

\usepackage{amsmath,amsfonts}
\usepackage{amsthm}

\usepackage[capitalise]{cleveref}  %
\usepackage{csquotes}

\usepackage[backend=biber,sorting=none,doi=false,isbn=false,url=true,
maxcitenames=2,mincitenames=1,style=ieee]{biblatex}
\usepackage{tikz}
\tikzset{every picture/.append style={font=\scriptsize}}  %
\newcommand\inputTikz[2][figures]{
    \let\pgfimageOld\pgfimage%
    \renewcommand{\pgfimage}[2][]{\pgfimageOld[##1]{#1/##2}}%
    \input{#1/#2}%
    \let\pgfimage\pgfimageOld%
}

\newtheorem{theorem}{Theorem}
\newtheorem{lemma}[theorem]{Lemma}
\newtheorem{proposition}[theorem]{Proposition}

\theoremstyle{definition}
\newtheorem{definition}{Definition}

\renewcommand{\vec}[1]{\boldsymbol{#1}}
\newcommand{\mat}[1]{\boldsymbol{#1}}
\newcommand*{\tran}{\mkern-2mu{}_{}^{\top}\mkern-3mu}

\DeclareMathAlphabet{\mathbfsf}{\encodingdefault}{\sfdefault}{bx}{sl}

\DeclareMathOperator*{\argmin}{arg\,min}

\usepackage{pifont}%

\usetikzlibrary{shapes,decorations,arrows,calc,arrows.meta,fit,positioning}
\tikzset{
    -Latex,auto,node distance =0.5 cm and 0.5 cm,semithick,
    state/.style ={ellipse, draw, minimum width = 0.5cm, minimum height = 0.5cm},
    box/.style={rectangle, draw, minimum width = 0.5cm, minimum height = 0.5cm},
    point/.style = {circle, draw, inner sep=0.04cm,fill,node contents={}},
    el/.style = {inner sep=2pt, align=left, sloped}
}

\usepackage[normalem]{ulem} %

\usepackage[most]{tcolorbox}

\begin{document}

\title{Controlling for Omitted Variable Bias\\ in Deep Neural Networks}

\author{%
  Manuel Pfeuffer\textsuperscript{1}\thanks{Corresponding author. E-mail: \texttt{manuel.pfeuffer@hu-berlin.de}.} \quad
  Roshan P.~Rane\textsuperscript{1,2,3} \quad
  Kerstin Ritter\textsuperscript{2,3} \quad
  Sonja Greven\textsuperscript{1} \\[0.4em]
  \textsuperscript{1}Humboldt-Universität zu Berlin, Berlin, Germany \\
  \textsuperscript{2}Hertie Institute for AI in Brain Health, Universität Tübingen, Tübingen, Germany \\
  \textsuperscript{3}Universität Tübingen, Tübingen, Germany
}

\maketitle

\begin{abstract}
Control variables are widely used in statistical modelling to account for omitted variable bias of known confounders.
However, they have largely been underexplored in deep learning.
This is surprising, given that deep learning models encode image-inferable covariates, such as demographic variables, into their predictions when these covariates are correlated with the outcome---a form of omitted variable bias referred to as `shortcut learning'.
While many existing confound-control or fairness methods try to restrict the correlation of such known covariates with model predictions, we show that this usually fails to obtain unbiased estimates and may remove important predictive information.
We therefore propose a control variable approach for deep learning models, based on generalised additive modelling of the effects of model inputs and covariates.
As flexible additive models can suffer from concurvity, we introduce an estimation procedure that refits the final layer of a pre-trained network to include covariate effects, using cross-fitting with ridge penalisation.
We show how these confounder-controlled direct effects can be decomposed into mediated and residual components and how to obtain confounder-adjusted predictions.
We verify our approach using simulated images, and demonstrate consistent estimation of effects.
In contrast, existing methods cannot handle continuous and multivariate confounders, require more data, and/or are unable to consistently estimate true effects.
We further apply our method to real neuroimaging data with experimentally induced confounding, where it recovers prediction performance to near the level of a model trained on unconfounded data.
Code is available at \url{https://github.com/mpff/ovb-dnn}.
\end{abstract}

\section{Introduction}
\label{sec:intro}

Deep neural networks (DNNs) are known to learn spurious associations due to covariates that are confounding or sensitive variables encoded in the input data, a phenomenon referred to as \emph{shortcut learning} \cite{geirhos2020ShortcutLearningDeep}.
Examples include medical image classifiers that use acquisition artefacts, such as rulers or markers, to predict disease labels \cite{zech2018VariableGeneralizationPerformance,}, or models that leverage demographic attributes such as sex or age when predicting neurological or psychiatric outcomes \cite{rane2024deeprepviz,klingenberg2023HigherPerformanceWomen}. 
While such associations may improve predictive performance on the training distribution, they typically fail to generalise out-of-sample and hinder meaningful interpretation and clinical applicability.

We argue that shortcut learning should be identified with \emph{omitted variable bias} (OVB) in classical regression analysis and addressed accordingly.
This bias arises when covariates $Z$ are correlated with both the DNN input $X$ and the outcome $Y$, causing variation in $Y$ associated with $Z$ to be incorrectly attributed to the DNN input variable $X$ through this joint correlation.
Assuming that these covariates are observed and available for modelling, OVB is addressed in classical regression models by the explicit inclusion of them as control variables (see e.g.\ \cite{angrist2009MostlyHarmlessEconometrics, wooldridge2010EconometricAnalysisCross}), thereby accounting for the association between $Z$ and $Y$ in the model.
Such variables could be demographic confounders, but also acquisition-related variables such as scanner type, imaging site, or visual artefacts.
However, despite their central role in classical statistics and econometrics, control variable techniques have, to our knowledge, seen little application in deep learning.

We propose incorporating covariates explicitly and additively into the final layer of DNNs, extending classical control-variable methods to deep learning models.
Our approach builds on work that connects the statistical theory of \emph{Generalised Additive Models} (GAMs) \cite{hastie1987GeneralizedAdditiveModels}, an interpretable and flexible regression framework, with modern deep learning through \emph{Neural Additive Models} (NAMs) \cite{agarwal2021NeuralAdditiveModels} and \emph{Semi-structured Networks} (SSNs) \cite{rugamer2023SemiStructuredDistributionalRegression}.
Because the final layer of a DNN is linear in its coefficients, it can be interpreted as an additive model operating on a learned feature representation.
Informed by this connection, we formalise shortcut learning as OVB arising within this additive structure.
This formulation provides a model-based framework for explaining and mitigating shortcut learning in DNNs, further grounding them in established statistical methodology.

In the biomedical applications that motivate our work, DNNs are increasingly used to test medical hypotheses.
Answering such questions requires modelling the \emph{data-generating process} (DGP) rather than maximising prediction accuracy alone \cite{rudin2019StopExplainingBlack}.
For example, when predicting alcohol misuse disorder from \emph{structural magnetic resonance imaging} (sMRI) \cite{rane2022StructuralDifferencesAdolescent}, the covariate $sex$ is correlated with misuse through social factors. 
Then, a DNN that can infer $sex$ from the scan attributes higher misuse probability to any male-looking brain, even when no disorder-related changes are present.
Incorporating covariates additively in the model instead separates the variation into an image component and a covariate component, so that the analysis can focus on the biomedically relevant image effect.
Much of this paper is devoted to how and when these components can be accurately identified.

In general, a covariate $Z$ may have a \emph{direct effect} on the outcome $Y$ as well as a \emph{mediated effect} on $Y$ through the input $X$.
If both pathways are present, we call $Z$ a \emph{confounder} for $X$ \cite{pearl2009Causality}.
We prove that OVB (and hence shortcut learning) arises whenever $Z$ has a direct effect on $Y$ and is correlated with $X$: it is this direct effect that gets attributed to $X$, while the mediated effect does not bias the estimate.
When $Z$ is a confounder, however, the mediated effect is additionally part of the $X$-effect --- which may or may not be desirable depending on the application.
We show that commonly used covariate control methods removing correlation of model predictions and confounders (e.g.\ \cite{madras2018LearningAdversariallyFair,ehsanadeli2019BiasResilientNeuralNetwork}) implicitly control for the mediated effects of $Z$, but usually give biased estimates.

Our method refits the last layer of a DNN to include control variables, either with or without an \emph{orthogonalisation} (secondary regression) of the last-layer features on the covariates.
The orthogonalisation allows estimating the effect of $X$ while excluding any mediated effects of $Z$, which can be relevant in fairness contexts, for example.
We restrict our approach to the last layer because \emph{concurvity} \cite{buja1986AdditivePrincipalComponents,ramsay2003EffectConcurvityGeneralized}, the non-linear analogue of \emph{multicollinearity}, prevents identifying covariate and network input effects in joint training when the sample size is small relative to the number of DNN parameters.
This issue is pronounced in DNNs, which can interpolate covariate effects in finite samples, as documented for deep additive models such as NAMs \cite{siems2023CurveYourEnthusiasm,zhang2025ChallengesInterpretabilityAdditive} and SSNs \cite{rugamer2023SemiStructuredDistributionalRegression,rugamer2023NewPHOrmulaImproved}.

For our refitting procedure we fit a partial linear model \cite{Speckman1988KernelSmoothingPLM} over features and covariates, while incorporating a ridge penalty to stabilise estimation in settings with small sample sizes.
As fitting the DNN backbone on the same sample used for refitting the last layer leads to biased estimates, we follow an approach proposed in \cite{chernozhukov2018DML} and cross-fit DNN backbone and last layer parameters over separate folds.
We provide a linear version of our approach suitable to regression tasks with real-valued output, as well as a generalised version based on the iteratively reweighted least squares (IRLS) algorithm \cite{nelder1972GeneralizedLinearModels} (suitable, e.g., for binary classification).
The ridge penalty is chosen using a validation data set, following the literature on elastic-net regularisation paths \cite{friedman2010RegularizationPathsGeneralized,tay2023ElasticNetRegularization}.

\section{Related Literature}
\label{sec:lit}

Classical statistical models explicitly include control variables to mitigate omitted variable bias.
In contrast, DNNs rarely include covariates in their architecture.
Notable exceptions are NAMs \cite{agarwal2021NeuralAdditiveModels} which additively model multiple inputs but were proposed for scalar features.
However, these features could also be, e.g., images and additional confounders as in \cite{polsterl2020WideDeepNeurala} and \cite{wolf2023DontPANICPrototypical}, who jointly train deep and covariate effects similar to NAMs.
As concurvity can make them unidentifiable and uninterpretable, especially when $X$ and $Z$ are correlated \cite{zhang2025ChallengesInterpretabilityAdditive},
SSNs \cite{rugamer2023SemiStructuredDistributionalRegression,rugamer2023NewPHOrmulaImproved} and concurvity penalties \cite{siems2023CurveYourEnthusiasm} were proposed to restore identifiability by orthogonalising or decorrelating the effects.
As we will see, they no longer estimate direct effects, but rather the \emph{residual effect} after mediated effects of the covariates have been removed (see \cref{sec:theo:med}).
Our method keeps the additive structure and allows for estimation of the effect of $X$ both with and without the mediated effects of $Z$.
Our methods are closely related to double/debiased machine learning \cite{chernozhukov2018DML} which estimates a low-dimensional parameter of interest while controlling for high-dimensional covariates.
However, we reverse the roles of target and control variables while adopting its cross-fitting strategy (see \cref{sec:met:cf}).

Unlike classical statistics, most confound control and fairness methods in deep learning try to remove covariate variation from model predictions. 
This can happen before, during or after training.
\emph{Dataset adjustment} methods preprocess or resample training data to remove correlation between covariates $Z$ and outcomes $Y$ \cite{calders2009BuildingClassifiersIndependency,kamiran2012DataPreprocessingTechniques,snoek2019HowControlConfounds,plecko2020FairDataAdaptation}.
In particular, deep feature reweighting \cite{kirichenko2023last} is closely related to our approach as it also retrains the last layer of a DNN backbone.
However, this becomes infeasible for many covariates, and continuous covariates are hard to balance without binning \cite{cochran_effectiveness_1968}.
\emph{Adversarial} methods control the correlation between covariates and predictions during training through penalised losses \cite{zemel2013LearningFairRepresentations,junpeikomiyama2018NonconvexOptimizationRegression,agarwal2018ReductionsApproachFair,cotter2019TrainingWellGeneralizingClassifiers,qingyuzhao2020TrainingConfounderfreeDeep,xianjingliu2021ProjectionwiseDisentanglingFair,enzotartaglione2021EnDEntanglingDisentangling,zhao2022InherentTradeoffsLearning,ventoanthony2022PenaltyApproachNormalizing,romanpogodin2022EfficientConditionallyInvariant}, whose hyperparameters set an opaque trade-off between bias and accuracy \cite{zliobaite2015RelationAccuracyFairness}, or strict constraints \cite{bashirsadeghi2019GlobalOptimaKernelized,lu2021MetadataNormalization,romanpogodin2022EfficientConditionallyInvariant}.
\emph{Environment}-based methods learn relationships that are invariant across datasets \cite{arjovsky2020InvariantRiskMinimization}, which requires multiple environments with stable causal or unconfounded structure \cite{rosenfeld2020RisksInvariantRisk}.
\emph{Post-hoc} methods remove correlation after training, e.g.\ by regressing covariates out of last-layer features \cite{t.weber2023UnreadingRacePurging}, or by erasing linear information about them \cite{norabelrose2023LEACEPerfectLinear,eliaschaibubneto2020CausalityawareCounterfactualConfounding}.
As we show in \cref{sec:whycorrfails}, decorrelating predictions from $Z$ implicitly controls for the mediated effects of $Z$, but may give biased estimates.
From a statistical perspective, such approaches suffer from model misspecification, because they do not explicitly model the effect of relevant covariates.
This can suffice for prediction tasks but falls short in high-stakes domains, where interpretability and generalisability are crucial \cite{rudin2019StopExplainingBlack}.

\section{Theoretical Framework}
\label{sec:theo}

We denote the random variables for the outcome by $Y$, for the covariates by $Z = (Z_1, \dots, Z_p) \in \mathbb{R}^p$ and for the network inputs by $X$, taking values in $\mathcal{S}$.
We denote by $\mathbb{E}[\cdot]$ the expectation with respect to their joint distribution and assume $\mathbb{E}[Y^2] < \infty$.
Our aim is to estimate the contribution of the network inputs $X$ to the conditional mean of $Y$, which we call the \emph{effect} of $X$ on $Y$.
We consider for now only a linear output function, so that the outcome $Y$ lies in $\mathbb{R}$, and cover the sigmoid or other generalised output functions in \cref{app:general}.
We describe such an effect by a deterministic function $f_X:\mathcal S\to\mathbb R$ or, alternatively, by the random variable $f_X(X)$ obtained by evaluation at the input $X$.

A typical assumption underlying many statistical models is \emph{additivity of effects} in the conditional expectation of $Y$ given $X$ and $Z$, so that
\begin{equation}
    \label{eq:ass-am}
    \mathbb{E}[Y\,|\, X, Z] = \beta_0 + f_{X}(X) + f_{Z}(Z),
\end{equation}
where $\beta_0 \in \mathbb{R}$ is the \emph{intercept}, $f_X(X) \in \mathcal{H}_X$ is the \emph{direct effect} of $X$, e.g.\ of an image, and $f_Z(Z) \in \mathcal{H}_Z$ is the \emph{direct effect} of $Z$.
To ensure identifiability of the intercept, we assume centred effects, and we require finite variance so that conditional means and mean squared errors are well defined.
Thus we assume that effects of $X$ lie in the space
\begin{equation}
  \label{eq:ass-cent}
  \mathcal H_X=\{g(X):\mathbb E[g(X)^2]<\infty,\ \mathbb E[g(X)]=0\},
\end{equation}
where $g$ ranges over measurable functions, and we define $\mathcal{H}_Z$ analogously.
It then follows that $\mathbb{E}[Y] = \beta_0$ by the \emph{law of iterated expectations} (LIE).

\subsection{Biased Estimation of $f_X$ due to Omitting Covariates $Z$}
\label{sec:theo:ovb}
\begin{figure}
\centering
\subfloat[
  DGP with image $X$ (blobs with intensities $v_1, v_2, v_3$), confounder $Z$ with effect strength $\beta_Z$, and outcome $Y$.
  \label{fig:bz:a}
]{
  \centering
  \includegraphics[width=0.28\linewidth]{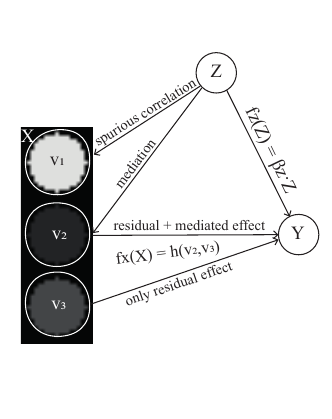}
}\hfill%
\subfloat[
  \emph{Mean squared prediction error} (MSPE) of $\hat f_X$ (left, with pixel-wise attribution) and $\hat f_X^\mathrm{re}$ (right) on a holdout test set, averaged over 50 model fits, with (top) and without (bottom) control variables.
  Bias-variance decompositions and results for $y, f_Z$ are given in \cref{app:sim_extra}.
  \label{fig:bz:b}
]{
  \centering
  \includegraphics[width=0.68\linewidth]{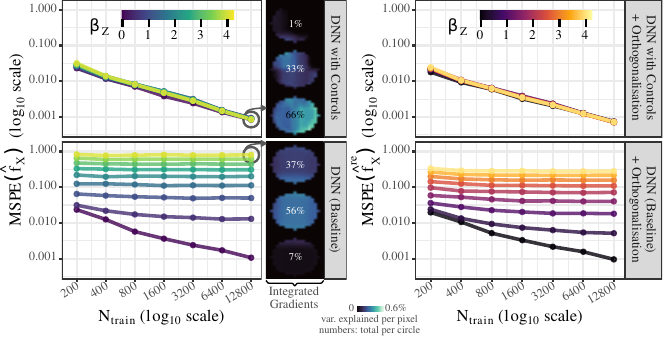}
}
\caption{Convergence for scalar outcomes $Y \in \mathbb{R}$ under increasing confounder strength $\beta_Z$.
Without control variables, estimates remain biased as $N_\mathrm{train}$ grows, and the bias increases with $\beta_Z$.
Our control variable method converges for all $\beta_Z$.
The images show the share of the variance of the estimated effect that Integrated Gradients \cite{sundararajan2017AxiomaticAttribution} attributes to each pixel for $\beta_Z = 4$ (circled), averaged over the 50 fits with reference image $\mathbb{E}[X]$.
Details on the simulation are given in \cref{sec:sim}.
\label{fig:bz}
}
\end{figure}

\noindent
Typically, only $X$ is used for predicting $Y$, so that $\mathbb{E}[Y \,|\, X]$ is estimated and not $\mathbb{E}[Y \,|\, X, Z]$.
Now, suppose the additive model in \cref{eq:ass-am} is in fact correct.
In this case, a model using only $X$ targets a biased version of the effect of $X$ on $Y$, due to \emph{omitted variable bias} (OVB) from omitting $Z$.
\begin{theorem}[OVB in Additive Models]
    \label{thm:ovb}
    Assume \cref{eq:ass-am} with \cref{eq:ass-cent}.
    Then, in a model omitting covariates $Z$, the minimiser of the \emph{mean squared error} (MSE),
    \begin{equation}
        \label{eq:gxstar}
        g_X^*(X) := \argmin_{g(X) \in \mathcal{H}_X} \mathbb{E}\bigl[(Y - \beta_0 - g(X))^2\bigr],
    \end{equation}
   is given by
    \begin{equation}
        \label{eq:ovb}
        g_X^*(X) = f_X(X) + \underbrace{\mathbb{E}[f_Z(Z)\,|\, X]}_{\text{OVB}}.
    \end{equation}
\end{theorem}
\vspace{-0.33em}
The proof can be found in \cref{app:ovb}.
The bias vanishes if and only if $\mathbb{E}[f_Z(Z) \,|\, X] = 0$, which holds in particular if $X$ and $Z$ are independent.
Otherwise, models trained solely on $X$ absorb the part of the direct effect $f_Z(Z)$ of the omitted covariate $Z$ that can be predicted from $X$.
This dependence on the function $f_Z$ in a DNN without control variables is demonstrated in \cref{fig:bz:b} (bottom left), while it disappears under our proposed control variable approach (\cref{fig:bz:b}, top left).

\subsection{Decomposing $f_X(X)$ into $Z$-Mediated and Residual Parts}
\label{sec:theo:med}
\noindent
Even when $Z$ is controlled for, its influence may still operate through $X$.
When $Z$ affects $X$, part of the effect $f_X(X)$ reflects variation in $X$ that is itself driven by $Z$, which is a mediated pathway rather than a spurious association.
In classical mediation analysis, such dependencies are quantified by \emph{path coefficients} obtained from linear regression \cite{baron1986ModeratorMediator}: the mediated effect of $Z$ on $Y$ is the product $\gamma_Z \beta_X$ of the coefficient $\gamma_Z$ linking $Z$ to $X$, obtained from estimating $\mathbb{E}[X\,|\, Z] = \gamma_0 + Z\gamma_Z$, and the coefficient $\beta_X$ linking $X$ to $Y$ given $Z$, obtained from estimating $\mathbb{E}[Y \,|\, X, Z] = \beta_0 + X\beta_X + Z\beta_Z$.
When $X$ is high-dimensional or its effect non-linear, as in DNNs, such coefficients are no longer available, and mediation may instead be defined on the level of outcomes \cite{pearl2001DirectIndirectEffects}.
In an additive model we can apply this idea to $f_X(X)$ directly and decompose it into a $Z$-mediated and a residual part.
\begin{definition}[Residual and $Z$-Mediated Part of $f_X(X)$]
\label{def:medeff}
Assume the additive model \cref{eq:ass-am} with \cref{eq:ass-cent}.
We define the \emph{$Z$-mediated} part of $f_X(X)$ as
\begin{equation}
  f_X^{\mathrm{me}}(Z) := \mathbb{E}[f_X(X) \,|\, Z],
\end{equation}
the component of the $X$-related effect that can be predicted from $Z$.
Then $f_X(X)$ decomposes into mediated and non-mediated parts $f_X(X) = f_X^{\mathrm{me}}(Z) + f_X^{\mathrm{re}}(X,Z)$, with
\begin{equation}
  f_X^{\mathrm{re}}(X,Z) := f_X(X) - \mathbb{E}[f_X(X) \,|\, Z],
\end{equation}
the \emph{residual effect} of $X$ given $Z$.
We define $f_Z^\mathrm{tot}(Z) := f_Z(Z) + f_X^\mathrm{me}(Z)$ as the \emph{total effect} of $Z$.
\end{definition}
For centred $X$ and $Z$ and linear relationships, the definition recovers the product of coefficients as the coefficient of $Z$, as then $\mathbb{E}[f_X(X) \,|\, Z] = \mathbb{E}[X \,|\, Z]\, \beta_X = Z \gamma_Z \beta_X$.
It generalises naturally to high-dimensional and non-linear settings by replacing path coefficients with the conditional expectations they describe.
Note that these definitions are purely statistical in nature, and a causal interpretation would require additional assumptions, such as $Z$ causally affecting $X$ and the absence of unmeasured confounding \cite{pearl2009Causality}, which is outside the scope of this paper.

\subsection{Why Correlation-based Confound Control Fails}
\label{sec:whycorrfails}

\noindent
Instead of including $Z$ as a control variable, many methods remove the correlation between predictions and covariates (see \cref{sec:lit}).
For example, they regress $Z$ out of predictions or last-layer features after training \cite{t.weber2023UnreadingRacePurging,norabelrose2023LEACEPerfectLinear}, or use adversarial penalties during training \cite{qingyuzhao2020TrainingConfounderfreeDeep,ehsanadeli2019BiasResilientNeuralNetwork}.
Formally, a function $g(X,Z)$ with finite variance is \emph{decorrelated} from $Z$ if $\mathbb{E}[g(X,Z) \,|\, Z] = \mathbb{E}[g(X,Z)]$, that is, if it is uncorrelated with every function of $Z$.
For centred functions, then even $\mathbb{E}[g(X,Z) \,|\, Z] = 0$.

\begin{proposition}[Limits of constraint-based covariate control]
    \label{prop:constraint}
    Assume \cref{eq:ass-am} with \cref{eq:ass-cent}.
    \begin{enumerate}
        \item (General) Every function $g(X,Z)$ with finite variance and $\mathbb{E}[g(X,Z) \,|\, Z] = 0$ satisfies
        \begin{equation}
            \label{eq:closer}
            \mathbb{E}\bigl[(g(X,Z) - f_X(X))^2\bigr] - \mathbb{E}\bigl[(g(X,Z) - f_X^\mathrm{re}(X,Z))^2\bigr] = \mathbb{E}\bigl[f_X^\mathrm{me}(Z)^2\bigr] \geq 0.
        \end{equation}
        Every such function is thus at least as close to $f_X^\mathrm{re}(X,Z)$ as to $f_X(X)$ in mean square.
        Among such decorrelated functions, $f_X^\mathrm{re}(X,Z)$ minimises $\mathbb{E}[(Y - \beta_0 - g(X,Z))^2]$.
        \item (Post hoc residualisation) Regressing $Z$ out of the minimiser $g_X^*(X)$ from \cref{thm:ovb}, that is, subtracting its conditional mean given $Z$, gives
        \begin{equation}
            \label{eq:posthoc}
            g_X^*(X) - \mathbb{E}[g_X^*(X) \,|\, Z]
            = f_X^\mathrm{re}(X,Z) + \underbrace{\mathbb{E}[f_Z(Z) \,|\, X] - \mathbb{E}\bigl[\mathbb{E}[f_Z(Z) \,|\, X] \,\big|\, Z\bigr]}_{\text{remaining OVB}}.
            \vspace{-0.66em}
        \end{equation}
        \item (Adversarial training) Among all $g(X) \in \mathcal{H}_X$ with $\mathbb{E}[g(X) \,|\, Z] = 0$, the minimiser of the MSE in \cref{eq:gxstar} is the one closest to $f_X^\mathrm{re}(X,Z)$ in mean square.
        It equals $f_X^\mathrm{re}(X,Z)$ if and only if $f_X^\mathrm{me}(Z)$ is also a function of $X$.
    \end{enumerate}
\end{proposition}

The proof can be found in \cref{app:constr-proof}.
\Cref{prop:constraint} shows that decorrelation changes the target from the direct effect $f_X(X)$ to the residual effect $f_X^\mathrm{re}(X,Z)$, whatever the method.
Decorrelated predictions therefore lose the mediated part $f_X^\mathrm{me}(Z)$ of $f_X(X)$.
The proposition further shows that common methods do not even recover $f_X^\mathrm{re}(X,Z)$ in general.
Post hoc residualisation of learned features (also called \emph{orthogonalisation} \cite{t.weber2023UnreadingRacePurging,rugamer2023NewPHOrmulaImproved}) does not remove the full OVB.
In \cref{fig:bz:b} (bottom right), its bias therefore grows with the strength of the covariate effect $\beta_Z$, although its predictions are uncorrelated with $Z$.
Adversarial training, in contrast, imposes the decorrelation during training, so the $Z$ information is no longer available to predict $Y$ and the model is not biased by OVB.
However, the resulting prediction tries to estimate $f_X^\mathrm{re}(X,Z)$ from $X$ alone, and it can therefore remove $f_X^\mathrm{me}(Z)$ from $f_X(X)$ only if this mediated part can be computed from $X$.

An example where both methods recover $f_X^\mathrm{re}(X,Z)$ is when covariates are visible artefacts, so that $Z = \varphi(X)$ for a deterministic function $\varphi$ of the image.
In dermoscopy, suspicious lesions are often marked, so we would like to control for a marks presence $Z$, which predicts the diagnosis \cite{zech2018VariableGeneralizationPerformance}.
Many covariates, however, are only partly inferable, such as age or sex in brain scans.
Both methods are then biased for $f_X^\mathrm{re}(X,Z)$ in general and both remove the mediated part $f_X^\mathrm{me}(Z)$, which is not always desirable: age-related brain differences may affect disease risk.
A control variable approach avoids both problems: it estimates $f_X(X)$ itself, and regressing $Z$ out of this estimate gives $f_X^\mathrm{re}(X,Z)$ without OVB (\cref{fig:bz:b}, top right), whereas decorrelation recovers $f_X^\mathrm{re}(X,Z)$ only in special cases.

\subsection{The Problem of Concurvity in the Estimation of $f_X$}
\label{sec:concurvity}

\noindent Estimation of $f_X$ is in general much harder than estimation of $f_X^\mathrm{re}$.
This difficulty mirrors variance inflation in linear models, where effects become harder to separate as $X$ and $Z$ become more correlated and are even unidentifiable under perfect multicollinearity.
In additive models, the non-linear analogue is called (approximate) \emph{concurvity} \cite{buja1986AdditivePrincipalComponents,buja1989LinearSmoothersAdditive}.
We distinguish two notions of concurvity.
\emph{Population concurvity} is a property of the joint distribution of $X$ and $Z$, and means that some non-zero $g(X) \in \mathcal{H}_X$ equals some $h(Z) \in \mathcal{H}_Z$ almost surely, i.e.\ $\mathcal{H}_X \cap \mathcal{H}_Z \neq \{0\}$ \cite{buja1989LinearSmoothersAdditive}.
Then, for any $c \in \mathbb{R}$, the decomposition
\begin{equation}
    \label{eq:shift}
    \mathbb{E}[Y \,|\, X, Z]
    = \beta_0
    + \underbrace{\bigl(f_X(X) + c\, g(X)\bigr)}_{\in\, \mathcal{H}_X}
    + \underbrace{\bigl(f_Z(Z) - c\, h(Z)\bigr)}_{\in\, \mathcal{H}_Z}
    \vspace{-0.33em}
\end{equation}
is again of the form \cref{eq:ass-am} with \cref{eq:ass-cent}, so that $f_X(X)$ and $f_Z(Z)$ are not identifiable.
For example, visible artefacts $Z$ lead to population concurvity, as $Z = \varphi(X)$ implies $\mathcal{H}_Z \subset \mathcal{H}_X$.
Without population concurvity, $f_X(X)$ and $f_Z(Z)$ are identifiable, but they become harder to separate the larger the dependence between $X$ and $Z$.
\begin{proposition}[Concurvity amplifies estimation error of $\hat f_X$]
    \label{prop:amplify}
    Assume \cref{eq:ass-am} with \cref{eq:ass-cent}, and let
    \begin{equation}
        \label{eq:rho}
        \rho := \sup\bigl\{\operatorname{corr}\bigl(g(X), h(Z)\bigr) : g(X) \in \mathcal{H}_X \setminus \{0\},\ h(Z) \in \mathcal{H}_Z \setminus \{0\}\bigr\}
    \end{equation}
    be the maximal correlation of $X$ and $Z$ \cite{renyi1959MeasuresDependence}.
    Let $\hat\beta_0$, $\hat f_X$ and $\hat f_Z$ be estimates, considered as deterministic functions given the training data, with $\hat f_X(X) \in \mathcal{H}_X$ and $\hat f_Z(Z) \in \mathcal{H}_Z$.
    Then
    \begin{equation}
        \label{eq:amplify}
        \mathbb{E}\bigl[(\hat f_X(X) - f_X(X))^2\bigr] \le (1 - \rho^2)^{-1}\, \mathbb{E}\bigl[(\hat Y - \mathbb{E}[Y \,|\, X,Z])^2\bigr],
    \end{equation}
    if $\rho < 1$, with $\hat Y := \hat \beta_0 + \hat f_X(X) + \hat f_Z(Z)$, and the same holds for $\hat f_Z(Z)$.
\end{proposition}
The proof can be found in \cref{app:amplify-proof}.
Here the right-hand side is the test error of the fitted model minus irreducible noise, so a model that predicts $Y$ well also estimates $f_X(X)$ well, up to the factor $(1 - \rho^2)^{-1/2}$.
The maximal correlation is zero if and only if $X$ and $Z$ are independent, and it is one under population concurvity.
As $\rho$ approaches one, a small error in the fitted values can become a large error in the effect estimates.
This is illustrated by the experiments in \cref{sec:sim} (\cref{fig:adversarial}, right).

\emph{Empirical concurvity} is instead a property of the model classes $\mathcal{F}_X, \mathcal{F}_Z$ and the sample: some $g \in \mathcal{F}_X$ and $h \in \mathcal{F}_Z$ take the same non-constant values on the training points.
It can occur even for independent $X$ and $Z$, for example if $\mathcal{F}_X$ can interpolate the training points, which arises easily for flexible classes, such as DNNs in NAMs \cite{siems2023CurveYourEnthusiasm,zhang2025ChallengesInterpretabilityAdditive} and SSNs \cite{rugamer2023SemiStructuredDistributionalRegression,rugamer2023NewPHOrmulaImproved}.
Under empirical concurvity, the training loss is invariant under shifts $\hat f_X + c\, g$, $\hat f_Z - c\, h$ of the estimates, $c \in \mathbb{R}$, as in \cref{eq:shift}.
The fit thus determines the fitted values, but not the two effects separately.

\begin{figure}[tb]
    \centering
    \includegraphics{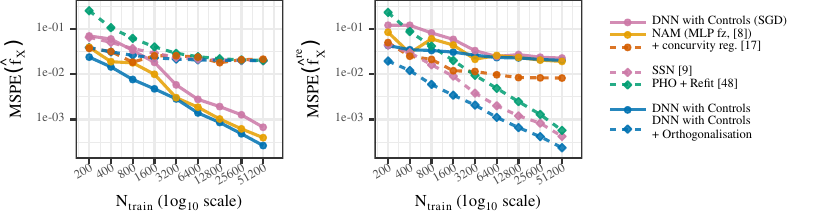}
    \caption{Convergence of $\hat f_X$ and $\hat f_X^\mathrm{re}$ across fitting strategies.
    Dashed lines indicate methods targeting $f_X^\mathrm{re}$.
    We fit a DNN with controls end to end with \emph{stochastic gradient descent} (SGD) and with our refit method (bottom), and for the NAM (using image valued inputs for $X$) we replace the linear covariate effect with a multilayer perceptron (MLP) \cite{agarwal2021NeuralAdditiveModels}.
    The NAM fit with the concurvity regulariser of \cite{siems2023CurveYourEnthusiasm} uses a penalty whose strength is chosen per $N_\mathrm{train}$ (\cref{app:concurvity-methods}).
    See \cref{fig:bz}a and \cref{sec:sim} for details on the simulation study and \cref{app:sim_extra} for further details on training and hyperparameter search.}
    \label{fig:concurvity}
\end{figure}

In contrast to population concurvity, identifiability may be recovered for increasing $n$, but the required sample size depends on the flexibility of $\mathcal{F}_X$.
In \cref{fig:concurvity}, we depict fitting behaviour for several methods under increasing sample size.
To recover $f_X$ already for small $n$, we propose to constrain the estimation to a much lower-dimensional, identifiable parameter, the last-layer coefficients of a pre-trained DNN.
This estimator shows the fastest convergence for $f_X$ and $f_X^\mathrm{re}$, respectively (dark green, light green), compared to other methods (see \cref{app:backbone-size,app:concurvity-methods} for details on the other methods).

\section{Methodology}
\label{sec:met}

\noindent
We constrain the estimation of the additive model to the last layer of a pre-trained DNN and include a ridge penalty to avoid overfitting, when the number of last-layer features $q \in \mathbb{N}$ is larger than $n \in \mathbb{N}$.
The \emph{backbone} of a DNN is a learned feature map $\phi: \mathcal{S} \rightarrow \mathbb{R}^q$, with network input space $\mathcal{S}$, obtained from minimising a loss $\ell(\vec y, \hat{\vec y})$ over $\vec y = (y_1, \dots, y_n)\tran$ and predicted outcomes $\vec{\hat y} = (\hat y_1, \dots, \hat y_n)\tran$ with $\hat{y}_i = \hat\beta_0 + \hat{\vec\phi}\tran_i \hat{\vec \beta}_X$ and $\hat{\vec\phi}_i = \hat\phi(X_i)$.
Applying $\hat\phi(\cdot)$ element-wise to inputs $\mat X \in \mathcal{S}^n$ yields the \emph{feature matrix} $\mat\Phi \in \mathbb{R}^{n \times q}$.
The refitted last layer consists of additive effects of covariates $\mat Z \in \mathbb{R}^{n \times p}$, the feature matrix $\mat \Phi$, and an intercept $\beta_0 \in \mathbb{R}$.
The outcome vector $\vec y$ may lie in $\mathbb{R}^n$, $\{0,1\}^n$, $\mathbb{N}_0^n$, or other spaces, depending on the assumed exponential family distribution of $Y$.
The resulting model corresponds to a \emph{partial (generalised) linear model} \cite{hastie1987GeneralizedAdditiveModels,rice1986ConvergenceRatesPartially,Speckman1988KernelSmoothingPLM}, where $f_X$ is linear in the learned features and $f_Z$ is a (possibly) flexible function.
We assume linear output functions throughout and provide a generalised version of our estimator for non-linear output functions (e.g.\ sigmoid) based on iteratively reweighted least squares in \cref{app:general}.

\subsection{Estimators in the Linear Case}
\label{sec:met:est}

\noindent Let $\phi := \hat\phi(X) \in \mathbb{R}^q$ denote the random feature vector given a pre-trained backbone and define the centred features $\tilde\phi := \phi - \mathbb{E}[\phi]$.
Then, by considering $f_X = \tilde\phi^\top \beta_X$ to be the last layer of a pre-trained DNN, the additive model \cref{eq:ass-am} can be written as
\begin{equation}
    \label{eq:plm}
    \mathbb{E}[Y \,|\, X, Z] = \beta_0 + \tilde\phi^\top \beta_X + f_Z(Z),
\end{equation}
where $\mathbb{E}[\tilde\phi] = 0$ ensures identifiability of $\beta_0$.
We consider estimation of $f_Z$ by a projection smoother $S_Z$, e.g.\ from a linear model or a spline basis, and define the centred smoother $\tilde S_Z = (I_n - \mathbf{1}_n \mathbf{1}_n^\top / n) S_Z$ and the centred residual maker $\tilde M_Z := I_n - \tilde S_Z$ matrices, where $\mathbf{1}_n$ is an $n$-vector of ones.
We estimate $f_X$ using ridge penalisation with $\lambda > 0$ to ensure a unique estimate of $f_X$ even for small $n$.

\begin{proposition}[Sample-Level Estimators for fixed $\lambda$]
    \label{prop:estimators}
    Let $\tilde y = y - \bar y \mathbf{1}_n$ and $\tilde\Phi = \Phi - \mathbf{1}_n \bar\phi^\top$, with $\bar\phi \in \mathbb{R}^q$ the vector of feature-wise means.
    Then, for a given $\lambda > 0$, the estimators for the additive model components in \cref{eq:plm} are given by
    \begin{align}
        \hat\beta_0 &= \frac{1}{n} \sum_{i=1}^n y_i = \bar y, \label{eq:beta0} \\
        \hat\beta_X(\lambda) &= \bigl(\tilde\Phi^\top \tilde M_Z \tilde\Phi + \lambda I_q\bigr)^{-1} \tilde\Phi^\top \tilde M_Z \tilde y, \label{eq:betax} \\
        \hat f_Z &= \tilde S_Z \bigl(\tilde y - \tilde\Phi \hat\beta_X\bigr), \label{eq:fz}
    \end{align}
    where $\lambda$ controls the strength of the ridge penalty in the estimation of $\hat\beta_X$.
\end{proposition}

The derivation can be found in \cref{app:est-proof}.
To select $\lambda$, we follow the regularisation path strategy of \cite{friedman2010RegularizationPathsGeneralized}, which is explained in detail in \cref{app:lambda-path}.
Given an estimate of $\hat f_X$, we can estimate the residual $X$-effect $\hat f_X^\mathrm{re}$ by post hoc orthogonalisation \cite{rugamer2023NewPHOrmulaImproved} as $\hat f_X^\mathrm{re}(\lambda) = \tilde M_Z \tilde\Phi \hat\beta_X(\lambda)$.

\subsection{Prediction Controlled for Covariates}
\label{sec:dis:ood}

\noindent
In many applications, obtaining predictions controlled for the covariates---and not just effect estimates $\hat{\vec f}_X$ controlled for covariates---is desirable.
We propose marginalising the model predictions over the distribution of $Z$ to achieve this, which mirrors the ``do''-operation from the causal inference literature \cite{pearl2009Causality}.
\begin{definition}[Marginalised Prediction]
\label{def:controlled-pred}
Let $p(z)$ denote the marginal distribution of $Z$.
For a new observation with network input $\mat X^*$, the \emph{marginalised prediction} is defined as
\begin{equation}
\label{eq:controlled-pred}
\hat y^*_{\mathrm{control}}
=
\int \hat{\mathbb E}[Y \mid X = \mat X^*, Z = z]\, p(z)\, \mathrm{d}z.
\end{equation}
\end{definition}
\vspace{-1em}
In practice, we approximate the integral by a sum over the empirical distribution of the covariates $Z$ in the training data.
In settings with dataset shift, $p^{\text{train}}(z)$ could also be replaced by a target distribution $p^{\text{test}}(z)$ to achieve better predictions.

\subsection{Consistent Estimation under Cross-fitting}
\label{sec:met:cf}

\noindent
\cref{prop:estimators} treats the feature matrix $\mat\Phi = (\hat\phi(X_1), \dots, \hat\phi(X_n))\tran \in \mathbb{R}^{n \times q}$ as a fixed design.
However, if $\hat\phi$ is trained on the same sample used to refit the last layer, it is no longer exogenous and estimators are biased \cite{chernozhukov2018DML,foster2023OrthogonalStatisticalLearning}.
Disjoint pre-training and refit samples restore exogeneity conditional on the pre-trained $\hat\phi$ and yield a consistent estimator under standard ridge regression conditions.

\begin{lemma}[Consistency of $\vec{\hat\beta}_X$ under sample splitting]
\label{lem:splitting}
Let $\mathcal{D}_\phi$ denote the pre-training sample of $\hat\phi$ and $\mathcal{D}$ the refit sample of size $n$, and suppose $\mathcal{D}_\phi \cap \mathcal{D} = \emptyset$.
Then conditional on the realised $\hat\phi$, for any penalty sequence with $\lambda_n / n \to 0$, and with certain smoother-quality assumptions on $\mat S_Z$,
\begin{equation*}
\vec{\hat\beta}_X(\lambda_n) \xrightarrow{p} \vec\beta_X^\circ,
\end{equation*}
where $\vec\beta_X^\circ$ is the minimum-$\ell_2$-norm population coefficient induced by the realised $\hat\phi$.
\end{lemma}

\noindent
The proof and more detailed assumptions are given in \cref{app:splitting-proof}.
\cref{lem:splitting} establishes consistency of the refitting procedure conditional on a given backbone, but not of the direct effect $f_X$ itself.
The quality of the refit also depends on how well $f_X$ can be approximated by a linear function of the learned features.
With $\hat f_X = \tilde\phi\tran \vec{\hat\beta}_X$ (up to centring), the estimation error decomposes into two parts,
\begin{equation}
    \label{eq:error-decomp}
    \hat f_X - f_X
    = \underbrace{\tilde\phi\tran \bigl(\vec{\hat\beta}_X - \vec\beta_X^\circ\bigr)}_{\text{refit error}}
    + \underbrace{\tilde\phi\tran \vec\beta_X^\circ - f_X}_{\text{backbone error}}.
\end{equation}
The refit error vanishes for increasing sample size under the conditions of \cref{lem:splitting}, but the backbone error depends on how well the true effect can be represented by a linear function of the pre-trained features.
However, by \cref{prop:amplify} it is at most $(1-\rho^2)^{-1/2}$ times the prediction error of the refitted model, so it is small whenever the pre-trained features predict $Y$ well together with $Z$.

Consistency of $\hat f_X$ for $f_X$ thus also depends on the quality of the pre-trained backbone, which in turn depends on the amount of data available for pre-training.
Sample splitting reduces this amount, as it recovers consistency at the cost of using half the training data to fit the DNN backbone $\hat\phi$.
This cost can be alleviated by using $K$-fold \emph{cross-fitting} proposed by \cite{chernozhukov2018DML}, where each fold is used once for the refit and otherwise for pre-training.

\begin{definition}[Cross-fitted ensemble model \cite{chernozhukov2018DML}]
\label{def:crossfit}
Let $\{\mathcal{D}_k\}_{k=1}^K$ partition the data into $K$ disjoint folds, and write $\mathcal{D}_{-k} := \bigcup_{l \neq k} \mathcal{D}_l$.
For each $k$, let $\hat\phi_{-k}$ be a backbone trained on $\mathcal{D}_{-k}$ and let $(\hat\beta_{0,k}, \vec{\hat\beta}_{X,k}, \hat f_{Z,k})$ denote the estimators of \cref{prop:estimators} computed on $\mathcal{D}_k$ using the features $\hat\phi_{-k}(\cdot)$ and the smoother $\mat{\tilde S}_{Z,k}$.
The cross-fitted predictor is
\begin{equation}
\label{eq:cf-ensemble}
\hat\eta(X, Z)
\;=\;
\frac{1}{K}\sum_{k=1}^{K}\Big( \hat\beta_{0,k} + \hat\phi_{-k}(X)\tran \vec{\hat\beta}_{X,k} + \hat f_{Z,k}(Z) \Big),
\end{equation}
with effect components $\hat f_X(\cdot) = K^{-1}\sum_k \hat\phi_{-k}(\cdot)\tran \vec{\hat\beta}_{X,k}$ and $\hat f_Z(\cdot) = K^{-1}\sum_k \hat f_{Z,k}(\cdot)$.
Since each $\hat f_{X,k}, \hat f_{Z,k}$ is centred only on fold $k$, we recentre $\hat f_X$ and $\hat f_Z$ over the full sample.
\end{definition}

\section{Simulation Study}
\label{sec:sim}

\noindent
To investigate convergence of estimates across different models and settings, we simulate images $X \in [0,1]^{20 \times 60}$, continuous confounders $Z \in [0,1]^p$ and an outcome $Y \in \{\mathbb{R}, \{0,1\}\}$ as follows:

\begin{figure}[!hbt]
    \vspace{-1.5em}
    \centering
    \begin{minipage}{0.58\textwidth}
        \centering
        \begin{align*}
            z_k &\sim U([0,1]) \quad \text{for $k=1,\dots,p$} \\
            \mat X &= h(\vec v_1, \vec v_2, v_3) \\ %
              & v_{1,k} \sim (1 - c_1) \cdot U([0,1]) + c_1 z_k \quad \text{for $k=1,\dots,p$}\\
              & v_{2,k} \sim (1 - c_2) \cdot U([0,1]) + c_2 z_k \quad \text{for $k=1,\dots,p$} \\
              & v_3 \sim U([0,1]) \\
            \eta &= \underbrace{\vec v_2^\top \vec \beta_2 + \beta_3 \cdot v_3}_{f_X} + \underbrace{\vec z^\top \vec\beta_Z}_{f_Z},
        \end{align*}
    \end{minipage}
    \hfill %
    \begin{minipage}{0.4\textwidth}
        \centering
        \includegraphics[width=\textwidth]{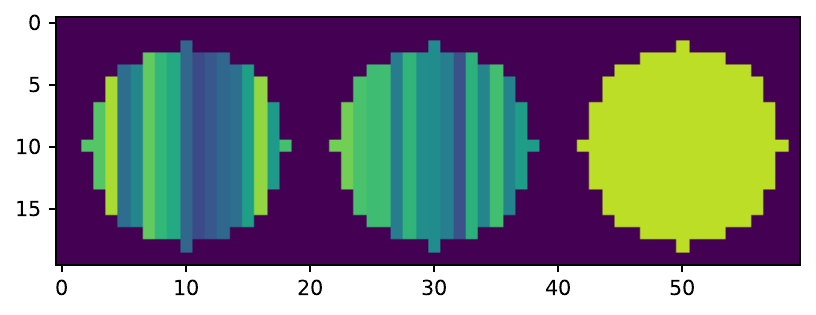}
        \vspace{-2em}
        \caption{Example image for $p=16$ confounders. Stripes have intensities $v_{1,k}$ (left), $v_{2,k}$ (centre), and $v_3$ (right).}
        \label{fig:sample}
    \end{minipage}
    \vspace{-1em}
\end{figure}

with $c_1, c_2 \in [0,1]$.
\cref{fig:sample} shows how $X$ is generated from $\vec v_{1,k}, \vec v_{2,k}$ and $v_3$.
For continuous outcomes, we draw $y = \eta + \epsilon$, $\epsilon \sim \mathcal{N}(0,1)$, and for binary outcomes, we use a logistic link with $p = \sigma(\eta - \mathbb{E}[\eta])$ and $y \sim \mathrm{Bernoulli}(p)$, where $\sigma$ denotes the sigmoid function.
Unless specified differently, we use $\beta_{2,k} = \beta_3 = \beta_Z = 1$, for $k=1,\dots,p$, $c_1 = 0.5$, $c_2 = 0.5$, $p=1$ as default parameters.
Additional figures and details on the backbone, the MSPE metric, and its bias-variance decomposition can be found in \cref{app:sim_extra}.

\Cref{fig:bz} summarises the main convergence behaviour of $\hat f_X$ under increasing $\beta_Z$: the proposed method (DNN with Controls, with and without orthogonalisation) is consistent as $N_\mathrm{train} \to \infty$, with both bias and variance of $\hat f_X$ decaying \emph{independently of} $\beta_Z$.
Methods without control variables (a baseline DNN and a post-hoc orthogonalised DNN) instead inherit an OVB that grows with $\beta_Z$, as derived in \cref{thm:ovb} and \cref{prop:constraint}.
The same results carry over to binary outcomes fit by IRLS (see \cref{app:gam}) and to non-linear covariate effects $f_Z$, e.g.\ estimated via a spline expansion of $Z$ (see \cref{app:sim_nonlinear}).
We further demonstrate better convergence than competing methods in \cref{fig:concurvity}.

\begin{figure}[!htb]
  \centering
  \includegraphics[width=\linewidth]{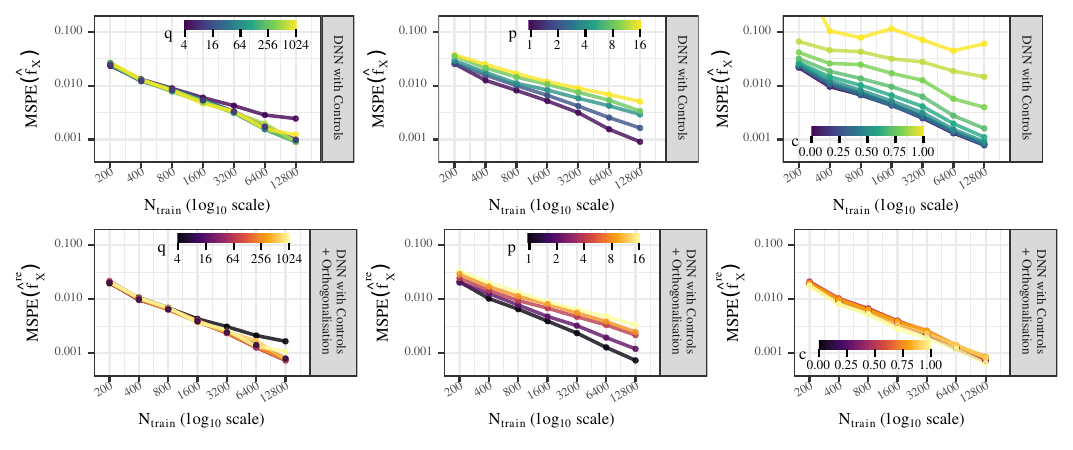}
  \caption{Convergence behaviour of $\hat f_X$ (top) and $\hat f_X^\mathrm{re}$ (bottom) in different adversarial settings for $y \in \mathbb{R}$ and $\beta_Z = 1$, with varying $q$ (left), $p$ (middle) and $c_1$ (right).}
  \label{fig:adversarial}
  \vspace{-1em}
\end{figure}

We additionally investigate three potential failure modes, shown in \cref{fig:adversarial}.
(left to right) \emph{Number of last-layer features $q$:} 
We see no difference in convergence behaviour over increasing $q$, except when it is chosen too small.
\emph{Number of confounders $p$:}
We see slower convergence of $\hat f_X$ and $\hat f_X^\mathrm{re}$ with increasing number of confounders.
Unlike many other methods relying on rebalancing training data, we can control for a large number of confounders; however, convergence will be slower.
\emph{Correlation between Image and Confounder $c_1$:}
We see slower convergence of $\hat f_X$ for increasing correlation of image and confounder, up to no convergence, when the confounder is perfectly encoded in $X$ ($c_1=1$, corresponding to $\rho = 1$).
This is the expected behaviour, as derived in \cref{sec:concurvity} on concurvity.
As orthogonalising $f_X$ with respect to the covariates $Z$ removes all correlation, the consistency of $\hat f_X^\mathrm{re}$ is not affected.

\section{Application to Neuroimaging Data with Synthetic Confounding}
\label{sec:app}

To evaluate our method in a high-dimensional real-data setting, we study the prediction of high alcohol consumption ($highalc$) from T1-weighted brain MRI \cite{rane2022StructuralDifferencesAdolescent}, in the UK Biobank~\cite{sudlow2015UKBiobankOpen} ($n_\mathrm{train} = 14{,}617$, $n_\mathrm{test} = 4{,}505$).
We introduce synthetic confounding by resampling the training data with replacement to follow a logistic DGP with $\beta_\mathrm{sex} = 2$ as the ``signal'' and $\beta_\mathrm{age} \in \{0, -2\}$ as the ``confounder''.
To prevent data leakage, we ensure that the original training dataset is first split into folds before resampling, and resample $5{,}000$ observations per fold and evaluate dependence on the confounder $age$ by resampling $2{,}500$ observations from a held-out test set with $\beta_\mathrm{age} = 0$, defining an \emph{unconfounded target} for measuring prediction performance.
This setting reflects the scenario in which the training distribution is confounded, yet unconfounded predictions are required at deployment.
Details can be found in \cref{app:ukbb-extra}.

The results are depicted in \cref{fig:ukbb-main} with tables and further results in the Appendix.
Every model uses a 3D ResNet-50~\cite{hara2018CanSpatiotemporalD} backbone before fine-tuning.
We evaluate four DNN models with $age$-control, that differ only in how the training fold is partitioned between backbone fitting and last-layer refit:
(i) \emph{No sample split} --- both stages share the full training fold ($5{,}000$ observations);
(ii) \emph{Sample split} --- backbone on one half ($2{,}500$ observations), refit on the disjoint other half;
(iii) \emph{Cross-fit (K=2)} --- two rotations of the sample split, ensembled;
(iv) \emph{Cross-fit (K=3)} --- three rotations.
AUC is measured on the resampled unconfounded test set across five outer cross-validation folds, with $age$-controlled predictions marginalised over the per-training-fold $age$ distribution (see \cref{def:controlled-pred}).

\begin{figure}[tb]
    \centering
    \includegraphics[width=\columnwidth]{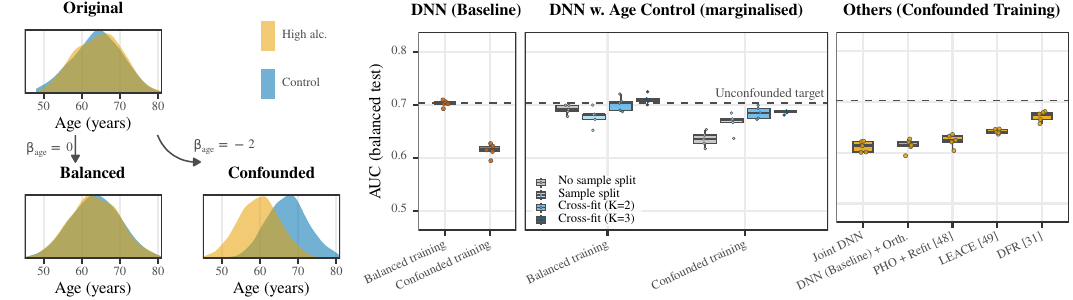}
    \vspace{-0.33em}
    \caption{(left) Visualisation of the resampling procedure. (right) Comparison of AUC over a held-out test set that was resampled so that $y$ is balanced with respect to $age$. $x$-axis indicates whether a model was trained on data that is balanced or $age$-confounded. 
    Other methods are described in \cref{app:ukbb-others} and are trained on $age$-confounded data. 
    \cref{tab:ukbb-others} in \cref{app:ukbb-extra} reports all AUCs.}
    \label{fig:ukbb-main}
\end{figure}

For the $age$-\emph{balanced} training setting ($\beta_\mathrm{age} = 0$), all five models agree closely (\cref{fig:ukbb-main}), however using sample splitting reduces performance, as effectively only half the sample is used to train the backbone and to refit the last layer coefficients.
Cross-fitting (K=2) recovers this drop in performance.
For the $age$-\emph{confounded} setting ($\beta_\mathrm{age} = -2$), the models diverge: the uncontrolled base DNN absorbs $age$ into its image features and its AUC on the balanced test set drops from $\approx 0.70$ to $\approx 0.62$.
Controlling for $age$ recovers most of this gap, but only when the backbone and refit step use disjoint data.
Cross-fitting with K=3 further reduces the variance across folds, compared to K=2.

We further compare our method with competing methods trained on the $age$-confounded data (\cref{fig:ukbb-main}, right, marginalised predictions).
A joint DNN receiving $age$ as an additional input cannot disentangle image and covariate effects.
Methods that decorrelate predictions or features from $age$ after training, such as orthogonalisation, PHO + Refit \cite{t.weber2023UnreadingRacePurging} and LEACE \cite{norabelrose2023LEACEPerfectLinear}, recover only a small part of the gap.
DFR \cite{kirichenko2023last}, which refits the last layer on group-balanced subsets, comes closest to our method; however, it requires discretising $age$ into groups, which would no longer work for the general setting of continuous outcomes and multivariate continuous confounders.
Overall, refitting the last layer with $age$ as a control variable recovers the largest part of the AUC lost to confounding.

\section{Limitations}
\label{sec:limitations}
We assume that covariates $Z$ are observed, which is not very restrictive in imaging cohorts, as demographic and acquisition-related variables such as scanner and site are routinely recorded.
Confounding that is never measured remains outside its scope.
The estimated image effect further depends on the expressiveness of the DNN backbone, and if the last-layer representation is not rich enough, it only approximates the direct effect within the span of the learned features.
When $Z$ is measured with noise, its effect is attenuated, and the bias of the image effect lies between that of a noise-free control and that of an uncontrolled DNN, and it persists asymptotically (\cref{app:noisyz}).
The additive decomposition does not itself interpret the deep image effect, but attribution methods such as Integrated Gradients \cite{sundararajan2017AxiomaticAttribution}, as used in \cref{fig:bz}, can complement it, as attributions computed for a confounder-controlled image effect inherit the covariate control.
Used for causal analysis, our method removes bias but the usual causal assumptions remain necessary for valid inference.

\section{Conclusion}
\label{sec:con}

We introduced a control variable approach for deep learning that estimates the direct image effect $f_X$ free of omitted variable bias by fine-tuning the last layer of a pre-trained network under cross-fitting with ridge penalisation, with an optional orthogonalisation step that recovers the residual effect $f_X^\mathrm{re}$.
Our theoretical results and simulations establish that last-layer refitting can recover the direct effect at lower sample sizes than existing methods, given that the covariate effects are correctly modelled.
Further we showed that correlation-based confound control can remove important predictive information in the form of mediated effects and may be biased in many settings.
Because our additive model separates $f_X$ from $f_Z$, predictions can be marginalised over the covariate distribution $p(Z)$ to control for $Z$, which also allows us to adapt to dataset shift by replacing $p^\mathrm{train}(z)$ with a target distribution $p^\mathrm{test}(z)$, a capability unavailable in models that merely decorrelate predictions from $Z$.
Together, these results bridge classical statistical additive modelling and deep learning, and open the door to interpretable, unbiased predictions in medical research.
A natural next step would therefore be multimodal learning, where the additive effects of different input modalities are identified, controlling for other modalities, even in small samples.
Another interesting direction would be extensions to include interaction effects of $X$ and $Z$, thereby relaxing the additivity assumption underlying our method.

\begin{ack}
We thank Georg Keilbar, Giulia Patané, Johannes Brandau, Maarten Jung, Marco Simnacher and Paulo Yanez for helpful discussions and feedback and Jeffrey Verhey for invaluable writing advice and for reviewing an early draft of the paper.
Funded by the Deutsche Forschungsgemeinschaft (DFG, German Research Foundation) – project number 459422098.
This research has been conducted using the UK Biobank resource under application number 33073.
The authors declare no competing interests.
\end{ack}

\printbibliography

\clearpage
\appendix
\section{Proofs and Derivations}
\label{proofs}

\subsection{Proof of \cref{thm:ovb}}
\label{app:ovb}

\begin{proof}
    Using the LIE and assumptions \cref{eq:ass-am,eq:ass-cent}, we can see that the conditional expectation of $\tilde Y := Y - \beta_0$ given $X$ is
    \begin{align}
         \mathbb{E}[\tilde Y \mid X]
         & \overset{\text{LIE}}{=} \mathbb{E}\bigl[\mathbb{E}[\tilde Y \mid X,Z]\mid X \bigr] \\
         & = \mathbb{E}[f_{X}(X) + f_Z(Z) \mid X] \\
         & = f_{X}(X) + \mathbb{E}[ f_Z(Z) \mid X].
    \end{align}
    It is well known that the conditional expectation is the best mean-square predictor and thus solves \cref{eq:gxstar} \cite[see e.g.][Sec.~2.7]{brockwell1991}.
    We write $m(X) := \mathbb{E}[\tilde Y \mid X]$ and follow the standard proof.
    
    First, $m(X)$ is centred and has finite variance so that $m(X) \in \mathcal{H}_X$, since
    \begin{equation}
        \mathbb E[m(X)] \overset{\text{LIE}}{=} \mathbb E[\tilde Y] = 0,
        \qquad
        \mathbb E[m(X)^2] = \mathbb E\bigl[\mathbb E[\tilde Y \mid X]^2\bigr] \overset{\text{Jensen}}{\le} \mathbb E\bigl[\mathbb E[\tilde Y^2 \mid X]\bigr] \overset{\text{LIE}}{=} \mathbb E[\tilde Y^2] < \infty,
    \end{equation}
    where $\mathbb E[\tilde Y] = 0$ follows from $\mathbb E[Y] = \beta_0$.
    Second, for every $h(X) \in \mathcal{H}_X$,
    \begin{equation}
        \mathbb E\bigl[(\tilde Y - m(X))\, h(X)\bigr] \overset{\text{LIE}}{=} \mathbb E\bigl[\mathbb E[\tilde Y - m(X) \mid X]\, h(X)\bigr] = \mathbb E\bigl[(m(X) - m(X))\, h(X)\bigr] = 0,
    \end{equation}
    where the product is integrable by the Cauchy--Schwarz inequality.
    Third, for any $g(X) \in \mathcal H_X$, the difference $h(X) = m(X) - g(X)$ also lies in $\mathcal{H}_X$.
    
    When considering the mean squared error of any $g(X) \in \mathcal{H}_X$ and adding and subtracting $m(X)$, the cross term then vanishes after multiplying out and the mean squared error decomposes as
    \begin{equation}
        \mathbb E\bigl[(\tilde Y - g(X))^2\bigr] = \mathbb E\bigl[(\tilde Y - m(X))^2\bigr] + \mathbb E\bigl[(m(X) - g(X))^2\bigr].
    \end{equation}
    Therefore it follows that, for any $g(X) \in \mathcal{H}_X$,
    \begin{equation}
        \mathbb E\bigl[(\tilde Y - g(X))^2\bigr] \geq \mathbb E\bigl[(\tilde Y - m(X))^2\bigr],
    \end{equation}
    with equality if and only if $g(X) = m(X)$ almost surely.
    Hence $g_X^*(X) = m(X) = f_X(X) + \mathbb{E}[f_Z(Z) \mid X]$ is the minimiser of \cref{eq:gxstar}.
\end{proof}

\subsection{Proof of \cref{prop:constraint}}
\label{app:constr-proof}

\begin{proof}
    Let $\tilde Y := Y - \beta_0$ and let
    \begin{equation*}
        \mathcal G := \bigl\{ g(X,Z) : \mathbb E[g(X,Z)^2] < \infty,\ \mathbb E[g(X,Z) \,|\, Z] = 0 \bigr\}
    \end{equation*}
    denote the set of decorrelated prediction functions, which is closed under linear combinations by the linearity of conditional expectations.
    We begin by establishing three auxiliary facts:
    First, every $g(X,Z) \in \mathcal G$ is uncorrelated with every square-integrable $h(Z)$, as
    \begin{equation}
        \label{eq:dec-orth}
            \mathbb E[g(X,Z)\, h(Z)] 
            \overset{\text{LIE}}{=} \mathbb E\bigl[\mathbb E[g(X,Z)\, h(Z) \,|\, Z]\bigr]
            = \mathbb E\bigl[\mathbb E[g(X,Z) \,|\, Z]\, h(Z)\bigr]
            = 0.
    \end{equation}
    Second, the residual effect $f_X^\mathrm{re}(X,Z) = f_X(X) - f_X^\mathrm{me}(Z)$ is an element of $\mathcal G$, as by \cref{def:medeff}
    \begin{equation}
        \mathbb E[f_X^\mathrm{re}(X,Z) \,|\, Z] = \mathbb E[f_X(X) \,|\, Z] - f_X^\mathrm{me}(Z) = 0,
    \end{equation}
    and it has finite variance, as $(a - b)^2 \le 2a^2 + 2b^2$ and Jensen's inequality give
    \begin{align}
        \mathbb E\bigl[f_X^\mathrm{re}(X,Z)^2\bigr]
        & = \mathbb E\Bigl[\bigl(f_X(X) - \mathbb E[f_X(X) \,|\, Z]\bigr)^2\Bigr] \\
        & \le 2\,\mathbb E\bigl[f_X(X)^2\bigr] + 2\,\mathbb E\bigl[\mathbb E[f_X(X) \,|\, Z]^2\bigr] \\
        & \overset{\text{Jensen}}{\le} 2\,\mathbb E\bigl[f_X(X)^2\bigr] + 2\,\mathbb E\bigl[\mathbb E[f_X(X)^2 \,|\, Z]\bigr] \\
        & \overset{\text{LIE}}{=} 4\,\mathbb E\bigl[f_X(X)^2\bigr] < \infty,
    \end{align}
    so that $f_X^\mathrm{re}(X,Z) \in \mathcal G$.
    Third, $\varepsilon := \tilde Y - \mathbb E[\tilde Y \,|\, X, Z]$ is uncorrelated with every square-integrable $h(X,Z)$, as
    \begin{equation}
        \label{eq:eps-orth}
            \mathbb E[\varepsilon\, h(X,Z)]
            \overset{\text{LIE}}{=} \mathbb E\bigl[\mathbb E[\varepsilon\, h(X,Z) \,|\, X,Z]\bigr]
            = \mathbb E\bigl[\mathbb E[\varepsilon \,|\, X,Z]\, h(X,Z)\bigr] = 0,
    \end{equation}
    where $\mathbb E[\varepsilon \,|\, X,Z] = \mathbb E[\tilde Y \,|\, X,Z] - \mathbb E[\tilde Y \,|\, X,Z] = 0$.

    For the general statement in 1., let $g(X,Z) \in \mathcal G$.
    Then $g(X,Z) - f_X^\mathrm{re}(X,Z) \in \mathcal G$, and \cref{eq:dec-orth} with $h(Z) = f_X^\mathrm{me}(Z)$ gives
    \begin{align}
        \mathbb E & \bigl[(g(X,Z) - f_X(X))^2\bigr]
         = \mathbb E\Bigl[\bigl((g(X,Z) - f_X^\mathrm{re}(X,Z)) - f_X^\mathrm{me}(Z)\bigr)^2\Bigr] \\
        & = \mathbb E\bigl[(g(X,Z) - f_X^\mathrm{re}(X,Z))^2\bigr] - 2\,\underbrace{\mathbb E\bigl[(g(X,Z) - f_X^\mathrm{re}(X,Z))\, f_X^\mathrm{me}(Z)\bigr]}_{=\,0} + \mathbb E\bigl[f_X^\mathrm{me}(Z)^2\bigr] \\
        & = \mathbb E\bigl[(g(X,Z) - f_X^\mathrm{re}(X,Z))^2\bigr] + \mathbb E\bigl[f_X^\mathrm{me}(Z)^2\bigr],
    \end{align}
    which is \cref{eq:closer}.
    For the MSE, \cref{eq:ass-am} and \cref{def:medeff} give $f_X(X) + f_Z(Z) = f_X^\mathrm{re}(X,Z) + f_Z^\mathrm{tot}(Z)$, so that
    \begin{equation}
        \tilde Y - g(X,Z) = \varepsilon + \bigl(f_X^\mathrm{re}(X,Z) - g(X,Z)\bigr) + f_Z^\mathrm{tot}(Z).
    \end{equation}
    The cross terms with $\varepsilon$ vanish by \cref{eq:eps-orth}, and the cross term of $f_X^\mathrm{re}(X,Z) - g(X,Z) \in \mathcal G$ with $f_Z^\mathrm{tot}(Z)$ vanishes by \cref{eq:dec-orth}, so that
    \begin{equation}
        \label{eq:mse-dec}
        \mathbb E\bigl[(\tilde Y - g(X,Z))^2\bigr] = \mathbb E[\varepsilon^2] + \mathbb E\bigl[(f_X^\mathrm{re}(X,Z) - g(X,Z))^2\bigr] + \mathbb E\bigl[f_Z^\mathrm{tot}(Z)^2\bigr].
    \end{equation}
    Only the second term depends on $g(X,Z)$, so the MSE is minimised if and only if $g(X,Z) = f_X^\mathrm{re}(X,Z)$ almost surely, which lies in $\mathcal G$.

    For post hoc residualisation in 2., we insert $g_X^*(X) = f_X(X) + \mathbb{E}[f_Z(Z) \,|\, X]$ from \cref{thm:ovb} and use the linearity of conditional expectations and \cref{def:medeff} to obtain
    \begin{align}
        g_X^*(X) - \mathbb{E}[g_X^*(X) \,|\, Z]
        & = f_X(X) - \mathbb{E}[f_X(X) \,|\, Z] + \mathbb{E}[f_Z(Z) \,|\, X] - \mathbb{E}\bigl[\mathbb{E}[f_Z(Z) \,|\, X] \,\big|\, Z\bigr] \\
        & = f_X^\mathrm{re}(X,Z) + \mathbb{E}[f_Z(Z) \,|\, X] - \mathbb{E}\bigl[\mathbb{E}[f_Z(Z) \,|\, X] \,\big|\, Z\bigr],
    \end{align}
    which is \cref{eq:posthoc}.

    For adversarial training in 3., let $\mathcal H_X^{\mathrm{dec}} := \mathcal H_X \cap \mathcal G$ be the set of decorrelated functions of $X$ alone.
    To apply the projection theorem, we first show that $\mathcal H_X^{\mathrm{dec}}$ is a closed subspace.
    For this, we rewrite the constraint $\mathbb E[g(X) \,|\, Z] = 0$ as orthogonality to every square-integrable $h(Z)$, that is, $\mathbb E[g(X)\, h(Z)] = 0$.
    The constraint implies orthogonality by \cref{eq:dec-orth}.
    Conversely, orthogonality to $h(Z) = \mathbb E[g(X) \,|\, Z]$, which is square-integrable by Jensen's inequality, gives
    \begin{equation}
        0 = \mathbb E\bigl[g(X)\, \mathbb E[g(X) \,|\, Z]\bigr] \overset{\text{LIE}}{=} \mathbb E\bigl[\mathbb E[g(X) \,|\, Z]^2\bigr],
    \end{equation}
    so that $\mathbb E[g(X) \,|\, Z] = 0$ almost surely.
    Hence $\mathcal H_X^{\mathrm{dec}}$ is the intersection of $\mathcal H_X$ with the orthogonal complement of the square-integrable functions of $Z$, and thus a closed subspace \cite[Prop.~2.3.1 and Problem~2.25]{brockwell1991}.

    We next identify the minimiser.
    Since $\mathcal H_X^{\mathrm{dec}} \subset \mathcal G$, the decomposition \cref{eq:mse-dec} holds for every $g(X) \in \mathcal H_X^{\mathrm{dec}}$.
    Only its second term depends on $g(X)$, so minimising the MSE over $\mathcal H_X^{\mathrm{dec}}$ is equivalent to minimising $\mathbb E\bigl[(f_X^\mathrm{re}(X,Z) - g(X))^2\bigr]$.
    By the projection theorem \cite[Thm.~2.3.1]{brockwell1991}, this minimiser exists, is unique, and is the orthogonal projection of $f_X^\mathrm{re}(X,Z)$ onto $\mathcal H_X^{\mathrm{dec}}$.

    Finally, the projection equals $f_X^\mathrm{re}(X,Z)$ if and only if $f_X^\mathrm{re}(X,Z) \in \mathcal H_X^{\mathrm{dec}}$ \cite[Prop.~2.3.2]{brockwell1991}.
    As $f_X^\mathrm{re}(X,Z) \in \mathcal G$ has finite variance and is centred by the LIE, this holds if and only if $f_X^\mathrm{re}(X,Z)$ is a function of $X$.
    As $f_X^\mathrm{re}(X,Z) = f_X(X) - f_X^\mathrm{me}(Z)$ and $f_X(X)$ is a function of $X$, this in turn holds if and only if $f_X^\mathrm{me}(Z)$ is a function of $X$, which is the claim in in 3.

    For visible artefacts $Z = \varphi(X)$, as in the example of \cref{sec:whycorrfails}, both $f_Z(Z) = f_Z(\varphi(X))$ and $f_X^\mathrm{me}(Z) = f_X^\mathrm{me}(\varphi(X))$ are functions of $X$.
    Hence $\mathbb{E}[f_Z(Z) \,|\, X] = f_Z(Z) = \mathbb{E}\bigl[\mathbb{E}[f_Z(Z) \,|\, X] \,\big|\, Z\bigr]$, so that the remaining OVB in \cref{eq:posthoc} vanishes.
    By 2.\ and 3.\, the population targets of both post hoc residualisation and adversarial training then equal $f_X^\mathrm{re}(X,Z)$.

    \end{proof}

\subsection{Proof of \cref{prop:amplify}}
\label{app:amplify-proof}

\begin{proof}
Let $u(X) := \hat f_X(X) - f_X(X) \in \mathcal{H}_X$, $v(Z) := \hat f_Z(Z) - f_Z(Z) \in \mathcal{H}_Z$ and $c := \hat\beta_0 - \beta_0$, so that $\hat y - \mathbb{E}[Y \,|\, X, Z] = c + u(X) + v(Z)$.
As $u(X)$ and $v(Z)$ are centred, the cross terms with $c$ vanish and
\begin{equation*}
    \mathbb E\bigl[(c + u(X) + v(Z))^2\bigr] = c^2 + \mathbb E\bigl[(u(X) + v(Z))^2\bigr] \ge \mathbb E\bigl[(u(X) + v(Z))^2\bigr].
\end{equation*}
If $u(X) = 0$ or $v(Z) = 0$, the statement follows directly, so let both be non-zero.
By definition of $\rho$, every non-zero $g(X) \in \mathcal H_X$ and $h(Z) \in \mathcal H_Z$ satisfy
\begin{equation*}
    \mathbb E[g(X)\, h(Z)] \le \rho \sqrt{\mathbb E[g(X)^2]\, \mathbb E[h(Z)^2]}.
\end{equation*}
Applying this to $g(X) = u(X)$ and $h(Z) = -v(Z)$ gives
\begin{equation*}
    -\mathbb E[u(X)\, v(Z)] \le \rho \sqrt{\mathbb E[u(X)^2]\, \mathbb E[v(Z)^2]}.
\end{equation*}
Hence
\begin{align*}
    \mathbb E\bigl[(u(X) + v(Z))^2\bigr]
    & = \mathbb E[u(X)^2] + 2\, \mathbb E[u(X)\, v(Z)] + \mathbb E[v(Z)^2] \\
    & \ge \mathbb E[u(X)^2] - 2\rho \sqrt{\mathbb E[u(X)^2]\, \mathbb E[v(Z)^2]} + \mathbb E[v(Z)^2] \\
    & = (1 - \rho^2)\, \mathbb E[u(X)^2] + \Bigl(\sqrt{\mathbb E[v(Z)^2]} - \rho \sqrt{\mathbb E[u(X)^2]}\Bigr)^2 \\
    & \ge (1 - \rho^2)\, \mathbb E[u(X)^2].
\end{align*}
Exchanging the roles of $u(X)$ and $v(Z)$ gives the statement for $\hat f_Z(Z)$.
\end{proof}

\subsection{Proof of \cref{prop:estimators}}
\label{app:est-proof}

\noindent
We first derive the so called normal equations based on conditional expectations for general functions $f_X(X)$ using backfitting \cite{buja1989LinearSmoothersAdditive}.
From constraining $f_X$ and $f_Z$ to be centred, it follows that
\begin{align}
    \mathbb{E}_Y[Y] & = \mathbb{E}_{X}[\mathbb{E}_{Y\,|\,X}[Y \,|\, X]] \\
        & = \mathbb{E}_{X}[\mathbb{E}_{Z\,|\,X}[ \mathbb{E}_{Y \,|\, X,Z}[Y \,|\, X, Z] \,|\, X]] \\
        & = \mathbb{E}_{X}[\mathbb{E}_{Z\,|\,X}[\beta_0 + f_X(X) + f_Z(Z) \,|\, X]] \\
        & = \beta_0 + \underbrace{\mathbb{E}_X[f_X(X)]}_{=0} + \underbrace{\mathbb{E}_X[\mathbb{E}_{Z \,|\, X}[f_Z(Z) \,|\, X]]}_{= \mathbb{E}_Z[f_Z(Z)] = 0} = \beta_0.
\end{align}
By considering the conditional expectations with respect to $X$ and $Z$ and setting $\tilde Y = Y - \beta_0$, we can derive normal equations for $f_X$ and $f_Z$, as
\begin{align}
    \mathbb{E}_{Y\,|\,X}[\tilde Y - f_Z(Z) \mid X] &= \label{eq:fx-pop}
        \mathbb{E}_{Z \,|\, X}[ \mathbb{E}_{Y \,|\, X, Z} [\tilde Y - f_Z(Z) \,|\, X, Z] \mid X]  \\
        &= \mathbb{E}_{Z \,|\, X}[ \mathbb{E}_{Y \,|\, X, Z} [\tilde Y \,|\, X, Z] - \mathbb{E}_{Y \,|\, X, Z} [f_Z(Z) \,|\, X, Z] \,|\, X] \\
        &= \mathbb{E}_{Z\,|\,X}[f_X(X) + f_Z(Z) - f_Z(Z) \mid X] \\
        &= \mathbb{E}_{Z\,|\,X}[f_X(X) \mid X] = f_X(X), \\
    \mathbb{E}_{Y\,|\,Z} [\tilde Y - f_X(X) \mid Z] &= \dots = f_Z(Z). \label{eq:fz-pop}
\end{align}

Starting with $\beta_0 = \mathbb{E}_Y[Y]$, we can replace the expectation with its sample analogue to obtain an estimator 
\begin{equation}
    \hat\beta_0 = \frac{1}{n} \sum_{i=1}^n y_i = \bar y.
\end{equation}
We use this to construct the centred vector of outcomes $\vec{\tilde y} = \vec y - \bar y \vec 1_n$.
We then consider $f_Z$ as an additive smoother, which means we estimate conditional expectations with respect to $Z$ by multiplication with a smoother matrix $\mat S_Z$ (see e.g.\ \cite{buja1989LinearSmoothersAdditive}).
To estimate a centred additive effect, we use a \emph{centred smoother matrix} defined as $\mat{\tilde S_Z} := (\mat I_n - \vec 1_n \vec 1_n^\top/n) \mat S_Z$, where $\mat I_n$ is the $n\times n$ identity matrix and $\vec 1_n$ is the $n \times 1$ vector of ones.
We can then replace the population equation for $f_Z$ (\cref{eq:fz-pop}) with its sample analogue based on the smoother matrix $\mat{\tilde S_Z}$ and obtain the estimator
\begin{equation}
    \vec{\hat f}_Z = \mat{\tilde S}_Z (\vec{\tilde y} - \mat{\tilde\Phi} \vec{\hat\beta}_X),
\end{equation}
which depends on the estimate $\vec{\hat\beta}_X$ and the \emph{centred feature matrix} $\mat{\tilde\Phi} \in \mathbb{R}^{n \times q}$.
To obtain $\vec{\hat\beta}_X$, we estimate the corresponding conditional expectation in \cref{eq:fx-pop} using least squares with ridge penalty, where we insert the formula for the estimator $\hat f_Z$, which also depends on $\vec{\hat\beta}_X$.
\begin{equation}
    \vec{\hat\beta}_X (\lambda) = \argmin_{\beta_X \in \mathbb{R}^q} \| \vec{\tilde y} - \mat{\tilde S}_Z (\vec{\tilde y} - \mat{\tilde\Phi} \vec\beta_X) -  \mat{\tilde\Phi} \vec\beta_X \|^2 + \lambda \| \vec\beta_X \|^2,
\end{equation}
where $\lambda > 0$ controls the strength of the ridge penalisation.
We can rewrite the objective function using the "residual maker" matrix $\mat{\tilde M}_Z := \mat I_n - \mat{\tilde S}_Z$ regressing out $Z$ from $Y$
\begin{equation}
    \vec{\hat\beta}_X (\lambda) = \argmin_{\beta_X \in \mathbb{R}^q}  \| \mat{\tilde M}_Z \vec{\tilde y} - \mat{\tilde M}_Z \mat{\tilde\Phi} \vec\beta_X \|^2 + \lambda \| \vec\beta_X \|^2.
\end{equation}
We assume $\mat{\tilde M}_Z$ to be idempotent with $\mat{\tilde M}_Z = \mat{\tilde M}_Z^2 = \mat{\tilde M}_Z^\top$, which is the case when $\mat{\tilde S}_Z$ is chosen as a projection type smoother (e.g.\ polynomials or basis splines) \cite{buja1989LinearSmoothersAdditive}.
The solution to the penalised least squares objective is then given by the well known ridge estimator \cite{hoerl1970RidgeRegressionBiaseda} for the $\mat{\tilde M}_Z$-residualised data
\begin{equation}
    \label{eq:fx:samp}
    \vec{\hat\beta}_X (\lambda) = \big( \mat{\tilde\Phi}^\top \mat{\tilde M}_Z \mat{\tilde\Phi} + \lambda \mat I_q \big)^{-1} \mat{\tilde\Phi}^\top \mat{\tilde M}_Z \vec{\tilde y}.
\end{equation}
An estimator for the direct effect can thus be obtained by fitting a ridge regression after regressing out the covariates from both the features and the outcome.
Here, $\lambda$ is a hyperparameter and usually chosen with leave-one-out cross-validation (see \cite{friedman2010RegularizationPathsGeneralized}) or by validation on a holdout dataset.
Except for the additional ridge penalty term in the inverse, this mirrors the Speckman estimator \cite{Speckman1988KernelSmoothingPLM} for the linear effect in a partial linear additive model.

\subsection{Proof of \cref{lem:splitting}}
\label{app:splitting-proof}

We consider the asymptotic setting to gain insight into the behaviour of our estimator.
We do not show consistency for the estimation of $\hat f_X$, which would depend on the quality of the pre-trained DNN backbone $\hat \phi$, as asymptotics of DNNs are out of scope for this work, but show consistency of $\hat{\vec\beta}_x(\lambda_n)$, conditional on some pre-trained $\hat \phi$, against the population minimum $\ell_2$ norm coefficient.
For this we need three sets of assumptions:
(a) assumptions on the correct specification of the model (b) assumptions on the smoother estimating $f_Z$ and (c) assumptions about the ridge penalty.
We assume
\begin{enumerate}
    \item[(a1)] $\vec y = \mat \Phi \vec \beta_X + \vec f_z + \vec\epsilon \quad \Leftrightarrow \quad y_i = \hat{\vec\phi}(\mat X_i)^\top \vec\beta_X + f_z(\vec z_i) + \epsilon_i,\quad i=1,\dots,n,$
    \item[(a2)] $\mathcal{D} = \{(y_i, \mat X_i, \vec z_i)\}$ is a sequence of i.i.d.\ random objects,
    \item[(a3)] $\mathbb E\big[(\hat{\vec\phi}(\mat X_i) - \mathbb E[\hat{\vec\phi}(\mat X_i)\mid \vec z_i]) (\hat{\vec\phi}(\mat X_i) - \mathbb E[\hat{\vec\phi}(\mat X_i)\mid \vec z_i])^\top\big] =: \mat\Sigma_{\phi\phi} $ is finite,
    \item[(a4)] $\mathbb{E}\big[(\hat{\vec\phi}(\mat X_i) - \mathbb E[\hat{\vec\phi}(\mat X_i)\mid \vec z_i]) \epsilon_i\big] = 0$.
\end{enumerate}
Note that conditional on $\mathcal{D}_\phi$ the backbone $\hat{\vec\phi}(\cdot)$ is a deterministic function, so from (a2) it immediately follows that $\{(y_i, \hat{\vec\phi}(\mat X_i), \vec z_i)\}$ is a sequence of i.i.d.\ random objects.
Further note that (a3) typically assumes a nonsingular matrix.
However, in the case of DNN features it can easily happen that some of them are redundant or linear combinations of the others, leading to singular (a3)---even after regressing out functions of $Z$.
Accordingly we investigate convergence against the minimum-$\ell_2$-norm coefficient $\vec\beta_x^\circ$ (which is unique) and not against some arbitrary population coefficient and it holds that $\mat \Phi \vec \beta_X = \mat \Phi \vec \beta_X^\circ$ for any $\vec\beta_X$ fulfilling (a1).
In particular, it holds that $\vec\beta_X^\circ = \mat\Phi^{+}\mat\Phi\vec\beta_X$ also for $n \to \infty$, where $(\cdot)^{+}$ indicates the Moore-Penrose pseudoinverse.

The proof uses a mixture of the standard proofs for ridge regression and partial linear models, while not assuming nonsingularity in (a3).
Without loss of generality, assume variables are centred and substitute (a1) into the estimator from \cref{prop:estimators}.
Then divide numerator and denominator by $n$ and define the rescaled penalty $\mu_n := \lambda_n / n$.
We get
\begin{align*}
\vec{\hat\beta}_X(\lambda_n)
  &= \big(\mat{\Phi}^\top \mat{ M}_Z \mat{\Phi} + \lambda_n \mat I_q\big)^{-1}\,\mat{\Phi}^\top \mat{ M}_Z \vec{ y} \\
  &= \Big(\tfrac{1}{n}\mat{\Phi}^\top \mat{ M}_Z \mat{\Phi} + \mu_n \mat I_q\Big)^{-1}\,\tfrac{1}{n}\mat{\Phi}^\top \mat{ M}_Z \vec{ y} \\
  &= \big(\mat{\hat A}_n + \mu_n \mat I_q\big)^{-1}\,\Big[\mat{\hat A}_n \vec\beta_X + \vec{\hat s}^{\,f\phi}_n + \vec{\hat s}^{\,\epsilon\phi}_n\Big],
\end{align*}
with
$$\mat{\hat A}_n := \tfrac{1}{n}\mat{\Phi}^\top \mat{ M}_Z \mat{\Phi},
\qquad
\vec{\hat s}^{\,f\phi}_n := \tfrac{1}{n}\mat{\Phi}^\top \mat{ M}_Z \vec{ f}_z,
\qquad
\vec{\hat s}^{\,\epsilon\phi}_n := \tfrac{1}{n}\mat{\Phi}^\top \mat{ M}_Z \vec{\epsilon}.$$
We then require the following limits in probability for the three terms:
$$\text{(b1)}\ \ \mat{\hat A}_n \xrightarrow{p} \mat\Sigma_{\phi\phi},
\qquad
\text{(b2)}\ \ \vec{\hat s}^{\,f\phi}_n \xrightarrow{p} \vec 0,
\qquad
\text{(b3)}\ \ \vec{\hat s}^{\,\epsilon\phi}_n \xrightarrow{p} \vec 0.$$
Conditions (b1) to (b3) require smoother-quality assumptions on $\mat{S}_Z$.
(b1) requires $\mat S_Z$ to be a consistent estimator of $\mathbb E[\hat\phi_j(\mat X)\mid \vec z]$, $j=1,\dots,q$. 
(b2) requires $\mat M_Z \vec f_Z \to 0$ so that the smoother is well-specified for the true target $\vec f_Z$, and (b3) then follows from (a4) once (b1) holds.
These are common assumptions in the partial linear model and we refer to \cite{Speckman1988KernelSmoothingPLM} for a rigorous treatment.
For example, regression splines with the basis dimension increasing (slowly) with $n$ deliver all three limits in probability \cite{newey1997ConvergenceRatesAsymptotic}.

Informally---disregarding limits for $\mu_n$ for now, applying (b1)-(b3) would yield 
\begin{equation*}
\beta_X(\lambda_n) \xrightarrow{p} \big(\mat{\Sigma}_{\Phi\Phi} + \mu_n \mat I_q\big)^{-1}\,\mat{\Sigma}_{\Phi\Phi} \vec\beta_X.
\end{equation*}
Given a decomposition $\mat \Sigma_{\Phi\Phi} = \mat U^\top \mat U$, where $\mat U$ is the population analogue to $\mat M_Z \mat \Phi / \sqrt{n}$, we can use the typical ridge regression assumption that
\begin{equation*}
\text{(c1)}\ \ \mu_n = \lambda_n/n \xrightarrow{n \to \infty} 0,
\end{equation*}
and apply the pseudoinverse limit
\begin{equation*}
(\mat U^\top \mat U + \mu_n \mat I_q)^{-1}\mat U^\top \xrightarrow{\mu_n \to 0} \mat U^{+},
\end{equation*}
with $\mat U^{+}$ the Moore--Penrose pseudoinverse of $\mat U$.
Taken together we can conclude that for $\mu_n \to 0$
\begin{equation*}
\vec{\hat\beta}_X(\lambda_n) \xrightarrow{p} \mat U^{+} \mat U \vec\beta_X.
\end{equation*}
The minimum-$\ell_2$-norm coefficient $\vec\beta_X^\circ$ is, by definition, that projection so that $\mat U^{+}\mat U\,\vec\beta_X = \vec\beta_X^\circ$ for any $\vec\beta_X$ satisfying (a1).
Then
\begin{equation*}
    \vec{\hat \beta}_X(\lambda_n) \xrightarrow{p} \vec\beta_X^\circ
\end{equation*}
i.e.\ the ridge estimator is consistent for the population minimum $\ell_2$-norm coefficient $\vec\beta_x^\circ$, conditional on the estimated backbone $\hat{\vec\phi}(\cdot)$.
\qed

\section{Generalisation to Non-linear Output Functions}
\label{app:general}

\subsection{Omitted Variable Bias for Non-linear Output Functions}
\label{app:gam}
\noindent In \cref{eq:ass-am} we assume that the conditional expectation is additively separable on the level of $Y$.
However, many common settings, such as disease status prediction, assume $Y$ to follow other distributions in the exponential family, such as a Bernoulli (e.g.\, binary labels) or Poisson (count data).
We can then assume the conditional expectation of $Y$ to be generalised additive with true regression function
\begin{equation}
    \label{eq:ass-gam}
    \mathbb{E}_{Y\,|\,X,Z}[Y\,|\, X, Z] = g^{-1}(\beta_0 + f_{X}(X) + f_Z(Z)),
\end{equation}
where $g : \mathbb{R} \rightarrow \mathbb{R}$ is a monotone \emph{link function} with $g^{-1}$ the \emph{response function}.
The response function $g^{-1}(\cdot)$ is typically called \emph{output function} in the machine learning literature and can be e.g.\, a sigmoid, exp or the identity function.
This covers models such as Logistic regression, Poisson regression, and others under the \emph{generalised additive model} (GAM) framework \cite{hastie1987GeneralizedAdditiveModels}.
In this setting we directly model the so-called \emph{additive predictor}
\begin{equation}
    \label{eq:gam-lp}
    \eta(X,Z) = \beta_0 + f_{X}(X) + f_Z(Z),
\end{equation}
which is linked to the conditional expectation of the output via the \emph{link function} by $\eta(X,Z) = g \left(\mathbb{E}_{Y\,|\,X,Z}[Y\,|\, X, Z]\right)$.

\begin{figure}[!hbt]
    \centering
    \includegraphics{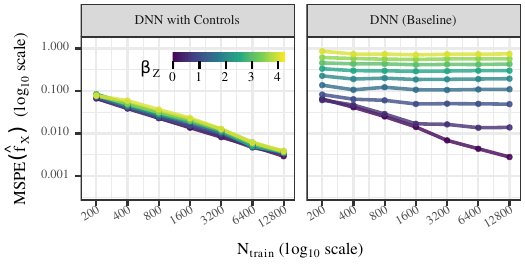}
    \caption{
        Convergence behaviour with binary outcome $Y \in \{0,1\}$ and sigmoid response function $g^{-1}(\cdot)$ under increasing strength $\beta_Z$ of $f_Z$ . 
        Also in this generalised case, controlling for covariates using our method leads to consistent estimation (left), here measured by $\mathrm{MSPE}(\hat f_X)$, while fitting a DNN without covariates (right) leads to inconsistent estimates with MSPE increasing with $\beta_Z$.
        See \cref{sec:sim} for details on the simulation study.
    }
    \label{fig:binary_bz}
\end{figure}

The preceding considerations translate to the generalised case as biases on the level of the additive predictor (e.g.\ on the level of the logits).
We can see this, by referring to usual Newton-Raphson-type algorithms for estimating GAMs based on \emph{iteratively reweighted least squares} (IRLS) (see e.g.\ \cite{wood2017GeneralizedAdditiveModels}), where the generalised model is iteratively approximated by a linear model using reweighted \emph{pseudo-data} as will be discussed in \cref{app:irls}.
The GAM is fitted iteratively by solving these reweighted linear models until convergence.
It follows that biases on the level of the linear pseudo-data model translate to the resulting solution, which is demonstrated in \cref{fig:binary_bz}.
We can see similar results to the linear case, as a standard DNN (right) shows bias increasing with $\beta_Z$ and a DNN with control variables (left) is unbiased.
As a side note, it is worth mentioning that in generalised models not including covariates $Z$ that are correlated with the outcome $Y$ can lead to underestimation of $f_X$ due to \emph{non-collapsibility} of the output function (see e.g.\ \cite{greenland1999ConfoundingCollapsibilityCausal}).

\subsection{Estimation for Generalised Output Functions}
\label{app:irls}
\noindent
In the non-linear case, we estimate the model using the standard iteratively reweighted least squares (IRLS) scheme for generalised additive models \cite{hastie1987GeneralizedAdditiveModels,wood2017GeneralizedAdditiveModels}.
Here, the non-linear estimation problem is approximated by a sequence of weighted least squares problems constructed from pseudo-responses and observation weights.

\begin{definition}[IRLS weights and pseudo-responses]
\label{def:irls}
Let $h = g^{-1}$ denote the inverse link.
Given current estimates $\hat\eta_i = \hat\beta_0 + \vec{\tilde\phi}_i^\top \vec{\hat \beta}_X + \hat f_Z(z_i)$ and $\hat\mu_i = h(\hat\eta_i)$, define weights
\begin{equation}
    w_i = \frac{h'(\hat\eta_i)^2}{V(\hat\mu_i)}
\end{equation}
and pseudo-responses
\begin{equation}
    y_i^* = \hat\eta_i + \frac{y_i - \hat\mu_i}{h'(\hat\eta_i)},
\end{equation}
where $V(\mu)$ denotes the variance function of the response distribution and $h'(\cdot)$ the derivative of the inverse link.
\end{definition}

\noindent
Define weighted centred variables 
$y_{w,i} = \sqrt{w_i}(y_i^* - \bar y_w),\;
\vec \phi_{w,i} = \sqrt{w_i}(\vec \phi_i - \vec{\bar\phi}_w),\;
\vec z_{w,i} = \sqrt{w_i}(\vec z_i - \vec{\bar z}_w)$
where $\bar y_w = (\vec w^\top \vec y^*)/(\vec w^\top \vec 1_n)$ and $\vec{\bar\phi}_w,\vec{\bar z}_w$ denote the corresponding weighted means.
Starting from an initial predictor $\hat\eta_i^{(0)} = \hat\beta_0^{(0)} = g(\bar y)$ and $\hat f_X \equiv \hat f_Z \equiv 0$, the weights and pseudo-responses from \cref{def:irls} are recomputed at each iteration.
Given these, $\vec{\hat\beta}_X$ and $\hat f_Z$ are the estimators from \cref{prop:estimators} computed on $\vec y_w$ and $\mat\Phi_w$, with $\mat{\tilde S}_Z$ replaced by the smoother $\mat S_{Z,w} = \mat Z_w(\mat Z_w^\top \mat Z_w)^{-}\mat Z_w^\top$ of the weighted, centred covariates.
As $\vec\phi$ and $\vec z$ are centred by unweighted means for identifiability, their weighted means do not vanish, and the intercept solving the weighted least-squares problem is
$\hat\beta_0 = \sum_i w_i \bigl(y_i^* - \vec{\tilde\phi}_i^\top \vec{\hat\beta}_X - \hat f_Z(\vec z_i)\bigr) / \sum_i w_i$.
The iteration stops at convergence of $\hat f_X$ and $\hat f_Z$.

For numerical conditioning of the ridge penalty, $\vec\phi$ and $\vec z$ are
centred and standardised to unit variance before the solve, for the identity link as for IRLS, so that
$\lambda$ penalises the coefficients of the standardised features.
$\hat{\vec\beta}_X,\hat{\vec\beta}_Z$ are de-standardised at the end, so that
$\hat f_X$ and $\hat f_Z$ are returned on the original centred-raw scale.
Convergence is monitored via the relative change of predicted $\vec{\hat f}_X \in \mathbb{R}^n$ and $\vec{\hat f}_Z \in \mathbb{R}^n$ between successive iterations.

For each candidate $\lambda \in \Lambda$, the IRLS algorithm is run to convergence before evaluating the validation loss $\mathcal{L}_{\mathrm{val}}(\lambda)$.
When $\lambda$ is chosen too small, it can happen that the algorithm fails to converge when $q$ is large.
This is known behaviour for regularised IWLS (see \cite{friedman2010RegularizationPathsGeneralized}) and usually not a problem.
It just indicates that larger $\lambda$ values should be used.

\section{Additional Considerations}

\subsection{Conditional Expectations and Smoother Matrices}
\label{app:smoothers}
\noindent 
Drawing on the literature on additive smoothers (e.g.\ \cite{buja1989LinearSmoothersAdditive}) we can consider a pre-trained DNN as a linear (in $\vec y$) smoother estimating conditional expectations with respect to network inputs $X$.
For example, we may estimate a DNN without covariates by smoothing $\mat\Phi \vec{\hat\beta_X} = \mat S_X \vec y$, with the DNN smoother matrix $\mat S_X$ given as the usual projection matrix $\mat S_X := \mat\Phi(\mat \Phi^\top \mat\Phi)^{-} \mat\Phi^\top$, where $(\cdot)^-$ denotes a generalised inverse.
Note that we do not need to assume that $\mat \Phi$ has full rank, as we can still uniquely identify the product $\mat\Phi \vec{\hat\beta}_X$ even if $\vec{\hat\beta}_X$ is not unique.
In settings where $q > n$ the DNN smoother matrix will interpolate the data, leading to poor generalisation. 
We can therefore define the \emph{ridge-penalised} smoother matrix $\mat S_X^\lambda = \mat\Phi(\mat \Phi^\top \mat\Phi + \lambda \mat I_q)^{-1} \mat\Phi^\top$ with $q \times q$ identity matrix $\mat I_q$ and penalty parameter $\lambda \geq 0$.
In contrast to $\mat S_X$, we want to make no further assumptions about the form of the covariate smoother matrix $\mat S_Z$.
Depending on whether $Z$ has linear or non-linear effects on $Y$, the smoother matrix $\mat S_Z$ could simply be given by $\mat S_Z = \mat Z(\mat Z^\top \mat Z)^{-1}\mat Z^\top$, corresponding to a linear model with $\mat Z \vec{\hat\beta_Z} = \mat S_Z \vec y$, or could be given by a projection into a space spanned by nonlinear basis functions $\mathcal{B}(\mat Z)$, such as cubic spline evaluations.

\subsection{On the Ridge-Regularisation Path and on Numerical Stability in IRLS Estimation}
\label{app:lambda-path}

\noindent
To select $\lambda$, we follow the regularisation path strategy proposed for penalised generalised linear models in \cite{friedman2010RegularizationPathsGeneralized}.
Specifically, we construct a decreasing grid $\Lambda = \{\lambda_{\max}, \dots, \lambda_{\min}\}$ of candidate values spaced logarithmically.
The final regularisation parameter $\hat\lambda$ is selected as
$\hat\lambda = \argmin_{\lambda \in \Lambda} \mathcal{L}_{\mathrm{val}}(\lambda)$
where $\mathcal{L}_{\mathrm{val}}(\lambda)$ denotes the validation loss obtained for a model fitted with regularisation parameter $\lambda$.
The resulting estimator $\vec{\hat\beta}_X(\hat\lambda)$ and the corresponding $\hat f_Z$ are then used as the final model.
If the optimal value occurs at the boundary of the candidate grid, the range $[\lambda_{\min},\lambda_{\max}]$ can be expanded and the procedure repeated in order to ensure that the optimal $\lambda$ lies within the explored path.

Following \cite{friedman2010RegularizationPathsGeneralized}, the regularisation path is constructed on a logarithmic grid $\Lambda = \{\lambda_{\max},\dots,\lambda_{\min}\}$. 
The upper bound $\lambda_{\max}$ corresponds to a strongly regularised model, while $\lambda_{\min}$ determines the weakest regularisation considered. 
We set $\lambda_{\min} = \epsilon \lambda_{\max}$ with $\epsilon = 10^{-3}$ for $q > n$ and $\epsilon = 10^{-6}$ for $q < n$.
For the identity link, $\lambda_{\max} = \|\mat\Phi\|_2^2$, where $\|\cdot\|_2$ denotes the spectral norm.
For the logit link, $\lambda_{\max}$ is computed using a weighted design motivated by the IRLS formulation of generalised linear models. 
Let $\mu$ denote the mean of the centred response, $V(\mu)$ the variance function, and $g'(\mu)$ the derivative of the link function. 
Weights are defined as
$w = g'(\mu)^2 / (V(\mu) + \varepsilon)$
with a small constant $\varepsilon$ for numerical stability, and
$\lambda_{\max} = 10 \|\mathrm{diag}(\sqrt{w})\,\mat\Phi\|_2^2$.
To avoid numerical instabilities in the binomial case, fitted probabilities $\hat\mu_i$ within $10^{-5}$ of $0$ or $1$ are set to exactly $0$ or $1$ and receive the weight $w_i = 10^{-5}$, so that they barely enter the weighted least-squares step.

\section{Additional Experimental Results}

\subsection{Details of the Simulation Study and Additional Figures}
\label{app:sim_extra}

We compare convergence behaviour of the estimators $\hat f \in \{\hat f_x, \hat f_x^\mathrm{re}\}$ of different models using the \emph{mean squared prediction error}
\begin{equation}\label{eq:mspe}
\text{MSPE}[\hat f] = \mathbb{E}_{(X,Z), \mathcal{D}} \left[ \left(f - \hat f_\mathcal{D} \right)^2 \right],
\end{equation}
where the expectation is taken over test observations $(y,X,Z) \in \mathcal{D}_\text{test} \sim p(\mathcal{D})$, training datasets are independently drawn from $p(\mathcal{D})$ and each training dataset yields a realisation $\hat f_\mathcal{D}$ of the estimator $\hat f$.
In practice, the expectations are replaced with sums over a test data of size $|\mathcal{D}_\text{test}| = 800$ and model fits estimated from $50$ independently drawn training data sets, each split into a training and validation part of equal size $N_\mathrm{train}$.

\begin{figure}[tb]
  \centering
  \includegraphics[width=\linewidth]{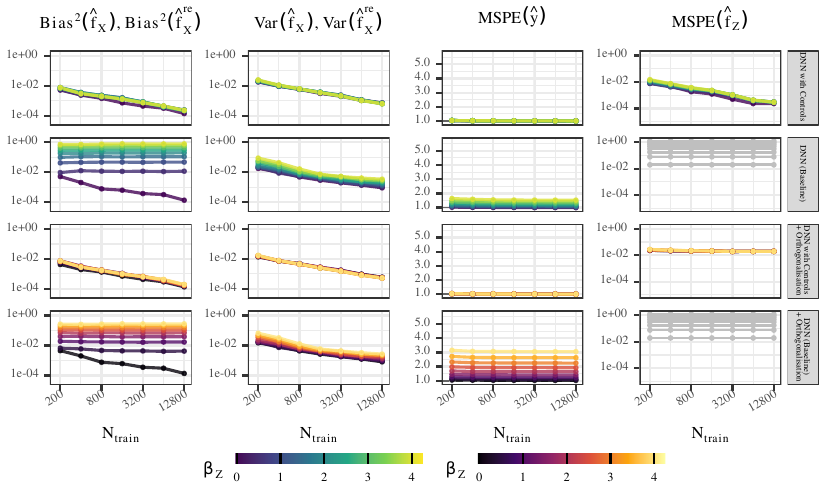}
  \caption{Bias--variance decomposition of $\hat f_X$ and MSPEs of $\hat y$ and $\hat f_Z$ in \cref{fig:bz}.
  It is clear that bias (first column) dominates the MSPE of models fit without covariates, while variance converges for all models as the sample size grows second column).
  For the orthogonalised models (bottom two rows), the first two columns show $\hat f_X^\mathrm{re}$.
  Only DNNs with controls predict $y$ well (MSPE near the noise floor) and only the DNN with covariates can disentangle $f_X$ and $f_Z$ effects.
  We can see that also $f_Z$ (here including the intercept $\beta_0$!) is consistently estimated when using a model with control variables, while models without controls have no estimate for $f_Z$ (greyed out line indicates $\beta_0$ only).}
  \label{fig:bz_appendix}
\end{figure}

To investigate the recovery of the true effects $f \in \{f_x, f_x^{re}\}$ we can further decompose the MSPE into bias and variance with $\text{MSPE}[\hat f] = \text{Bias}^2[\hat f] + \text{Var}[\hat f]$.
We calculate bias and variance via
\begin{align}
    \text{Bias}^2[\hat f] &= \mathbb{E}_{(X,Z)}\left[ \left( \mathbb{E}_\mathcal{D}[\hat f_\mathcal{D}] - f \right)^2 \right], \\
    \text{Var}[\hat f] &= \mathbb{E}_{(X,Z),\mathcal{D}} \left[ \left(\hat f_\mathcal{D} - \mathbb{E}_\mathcal{D}[\hat f_\mathcal{D}] \right)^2 \right].
\end{align}
and display additional MSPE decomposition results and the MSPEs of $\hat y$ and $\hat f_Z$ in \cref{fig:bz_appendix}.

For binary outcomes, the centring of $\eta$ ensures that the linear predictor does not saturate the sigmoid, so that both classes occur with non-negligible probability.
For evaluation, we centre $f_X$ and $f_Z$ around their population mean to align with our model definition.
We calculate the population $f_X^\mathrm{re}$ by removing $\mathbb{E}[\vec v_2^\top \vec \beta_2 \,|\, Z = \vec z] = c_2 \cdot \vec z^\top \vec\beta_2$ (the $Z$-mediated part of $v_2$) from $f_X$.

The last layer feature representation of the used DNN backbone has size $q$ which is set to 32 as default.
The backbone consists of two 2D convolutional layers \cite{lecun1998GradientbasedLearningApplied,krizhevsky2012ImageNetClassificationDeep} (channels: $1 \rightarrow 16, 16 \rightarrow 32$; kernel size 3, padding 1), an adaptive average 2D pooling \cite{nagi2011MaxpoolingConvolutionalNeural} (output: $4 \times 4$) and a fully connected layer of size $32 \cdot 4 \cdot 4 \rightarrow q$.
$f_Z$ and $f_X$ are (mean zero) linear functions of $Z$ and the last layer features, respectively.
Our method is fit with the $2$-fold cross-fit ensemble (\cref{def:crossfit}) throughout the simulation studies.

We optimise all backbones with Adam, learning rate $3\!\times\!10^{-3}$, weight decay $10^{-5}$, batch size $200$.
Models are fit using early stopping with patience $6$, \texttt{ReduceLROnPlateau} scheduler (mode \texttt{min}, patience $5$, factor $0.5$) anneals the learning rate.
The post-hoc step is solved with a glmnet-style ridge path of $100$ log-spaced $\lambda$ candidates with adaptive end-point expansion, and $\lambda$ selected via prediction loss on the validation half.
The IRLS algorithm is run up to a coefficient-change tolerance of $10^{-2}$.
Unless stated otherwise we sweep sample size over $n\!\in\!\{400, 800, 1600, 3200, 6400, 12800, 25600\}$ and at simulation defaults $\beta_z=b_2=b_3=1$, $c_1=c_2=0.5$, $\sigma_y=1$.
Competing methods are fit on the full training sample.
Learning rate and weight decay are shared by all methods in the linear, binary, nonlinear, and noisy-covariate studies.

For the comparison to other methods in \cref{fig:concurvity}, hyperparameters are instead searched per method and per sample size.
We select the learning rate and weight decay from the grid $\{3\times10^{-4}, 10^{-3}, 3\times10^{-3}, 10^{-2}\} \times \{10^{-5}, 10^{-4}\}$ at three sample sizes $n \in \{400, 6400, 102400\}$, and for the backbone-size sweep at nine pairs $(n, q) \in \{400, 3200, 25600\} \times \{4, 64, 1024\}$.

\subsection{Overview of Methods Compared in \cref{fig:concurvity}}
\label{app:concurvity-methods}

\begin{figure}[tb]
    \centering
    \includegraphics[width=\linewidth]{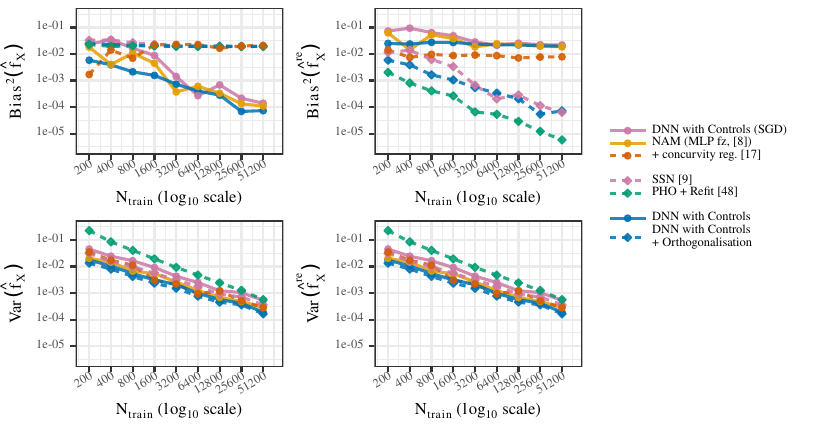}
    \caption{$\mathrm{Bias}^2$ (top) and $\mathrm{Var}$ (bottom) of $\hat f_X$ (left) and $\hat f_X^\mathrm{re}$ (right) for the methods of \cref{fig:concurvity}, computed over the simulation runs at each test point and averaged over the test points, so that $\mathrm{MSPE} = \mathrm{Bias}^2 + \mathrm{Var}$.}
    \label{fig:concurvity_bias_var}
\end{figure}

All competing methods in \cref{fig:concurvity} use the same backbone as our method, are fit on the full training sample of size $N_\mathrm{train}$ with the squared error loss, and report as $\hat f_X$ the image part of their predictor, centred on the training sample.
Learning rate and weight decay are searched per method and per $N_\mathrm{train}$ (\cref{app:sim_extra}).
\Cref{fig:concurvity_bias_var} shows the results of the $\mathrm{MSPE}$ in \cref{fig:concurvity} decomposed into $\mathrm{Bias}^2$ and $\mathrm{Var}$.

\paragraph{Fitting routines.}
The methods in both experiments differ in how the backbone is trained and on which data the last layer is fit.
(i) Joint fit with $Z$: DNN with Controls (SGD), NAM (MLP $f_Z$) and SSN train the backbone together with a model for $Z$ on the full training sample.
(ii) Training with a penalty: the concurvity regulariser trains the backbone (using MLP $f_Z$) on the full training sample with a penalty on the dependence between $\hat f_Z$ and $\hat f_X$.
(iii) Post-hoc correction: SSN and PHO + Refit start from a network fit on the full training sample, DNN with Controls (SGD) for SSN and DNN (Baseline) for PHO + Refit, and correct its predictor or its features on the same sample. PHO + Refit then refits the last layer.
(iv) Last-layer refit on held-out data: the DNN with Controls (sample split, cross fit) trains the backbone without $Z$ on one part of the training sample and refit the last layer on the other part.

\paragraph{DNN with Controls (SGD) and NAM (MLP $f_Z$) \cite{agarwal2021NeuralAdditiveModels}.}
DNN with Controls (SGD) is the additive model $\eta = \beta_0 + f_X(X) + \beta_Z Z$, with $f_X$ the backbone followed by a linear layer, fit end to end with Adam.
NAM (MLP $f_Z$) is the same model with $\beta_Z Z$ replaced by an MLP $f_Z(Z)$ with two hidden layers of width $32$ and ReLU activations.

\paragraph{SSN \cite{rugamer2023SemiStructuredDistributionalRegression,rugamer2023NewPHOrmulaImproved}.}
The fitted DNN with Controls (SGD), followed by post-hoc orthogonalisation: $\hat b = \arg\min_b \lVert \hat f_X - Z b \rVert^2$ on the training sample, and $\hat f_X - Z \hat b$ replaces $\hat f_X$, while $Z \hat b$ is added to $\hat f_Z$, so that $\hat\eta$ is unchanged.
SSN therefore estimates $f_X^\mathrm{re}$.

\paragraph{PHO + Refit \cite{t.weber2023UnreadingRacePurging}.}
The DNN (Basline), followed by post-hoc orthogonalisation.
The last layer is refit by least squares without a penalty, which we assume is the reason for the high variance / low bias result in \cref{fig:concurvity_bias_var}. Our method instead uses crossfitting and ridge penalisation to stabilise estimation in any setting.
We excluded one run at $N_\mathrm{train} = 400$ where the refit failed.

\paragraph{NAM with concurvity regulariser \cite{siems2023CurveYourEnthusiasm}.}
NAM (MLP $f_Z$) with $\kappa \, \lvert \mathrm{Corr}(\hat f_X(X), \hat f_Z(Z)) \rvert$, computed on each minibatch, added to the loss.
As in \cite{siems2023CurveYourEnthusiasm}, the penalty is inactive during the first $5\%$ of epochs.
The learning-rate schedule and early stopping on the validation squared error start when the penalty becomes active.
Learning rate and weight decay are those of NAM (MLP $f_Z$).
For each $N_\mathrm{train}$, $\kappa$ is chosen from five values ($\{10^{-3}, 10^{-2}, 10^{-1}, 1, 10\}$) by the lowest mean $\mathrm{MSPE}(\hat f_X^\mathrm{re})$ over five simulated data sets that are disjoint from those of the study, evaluated on a test set other than the study's.
The regulariser removes the dependence on $Z$ from $f_X$, so its target is $f_X^\mathrm{re}$.
As the selection uses the true $f_X^\mathrm{re}$ (oracle), the method is shown at its best strength in hindsight.

\subsection{Nonlinear Covariate Effect}
\label{app:sim_nonlinear}

To verify that the framework extends to non-linear covariate effects, we repeat the $\beta_Z$ sweep with the linear $f_Z$ replaced by $f_Z(Z) = \beta_Z \cdot \sin\!\left(2\pi\,(Z - 0.5)\right)$, %
keeping the image DGP and all other parameters unchanged.
$f_Z$ is now estimated as a cubic B-spline regression: we use the basis evaluations $\mathcal{B}(Z) \in \mathbb{R}^{n \times 9}$ ($5$ inner knots evenly spaced over $[0,1]$) as control variables and use cross-fitting with $K=2$.

\begin{figure}[tb]
    \centering
    \includegraphics[width=\linewidth]{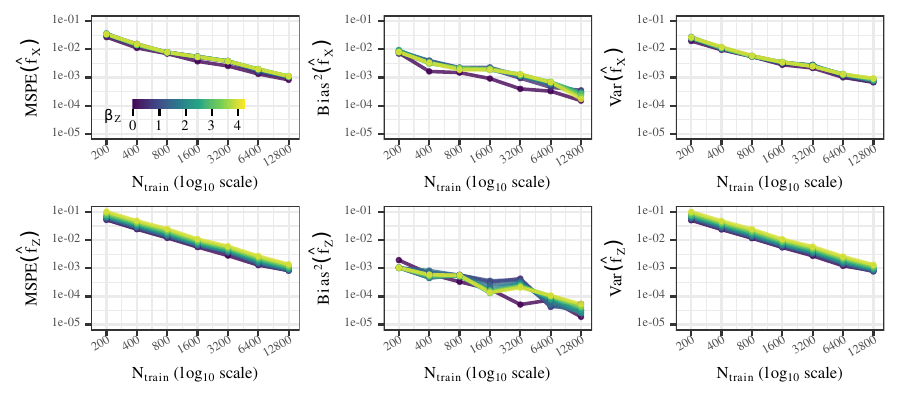}
    \caption{Convergence behaviour under non-linear $f_Z$ for the DNN w.\ Controls estimator with $k=2$ cross-fitting and a cubic B-spline basis on $Z$.
    \emph{Top}: MSPE and bias-variance decomposition of $\hat f_X$. 
    \emph{Bottom}: MSPE and bias-variance decomposition of $\hat f_Z$.
    Curves are nearly independent of $\beta_Z$.}
    \label{fig:nonlinear_fz}
\end{figure}

\Cref{fig:nonlinear_fz} confirms that
(i) the bias and variance of $\hat f_X$ remain essentially independent of $\beta_Z$, so the spline absorbs the non-linear covariate effect without leaking into the image-effect estimate, and
(ii) $\mathrm{Var}(\hat f_Z)$ scales slightly with $\beta_Z$ while $\mathrm{Bias}^2(\hat f_Z)$ stays well below the variance, so the spline regression for $f_Z$ is also consistent across the full $\beta_Z$ range.

\begin{figure}[tb]
    \centering
    \includegraphics[width=\linewidth]{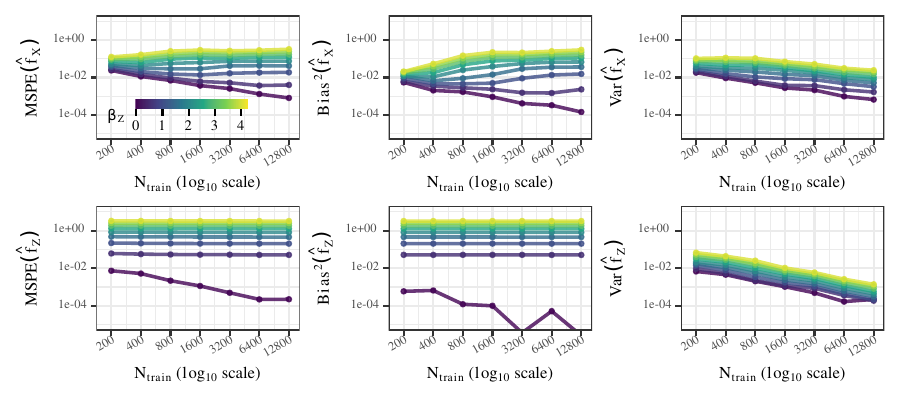}
    \caption{Convergence behaviour for non-linear $f_Z$ under model misspecification.
    Here we control for only linear effects of $Z$, even though the true $f_Z$ is non-linear.
    \emph{Top}: MSPE and bias-variance decomposition of $\hat f_X$.
    \emph{Bottom}: MSPE and bias-variance decomposition of $\hat f_Z$.}
    \label{fig:nonlinear_fz_misspec}
\end{figure}

In \cref{fig:nonlinear_fz_misspec} we see however that controlling for $Z$ only linearly does not suffice in this setting.
The part of $f_Z$ that is not linear in $Z$ is then an omitted variable: $\mathrm{Bias}^2(\hat f_Z)$ stays constant in $N_\mathrm{train}$ and $\mathrm{Bias}^2(\hat f_X)$ increases with $N_\mathrm{train}$ for all $\beta_Z > 0$.
As $X$ depends on $Z$, the backbone can learn this part of $f_Z$ from $X$, and it does so better with more data.
The control variables therefore have to be flexible enough to represent $f_Z$, as with the spline basis in \cref{fig:nonlinear_fz}.

\subsection{Convergence of Refit and Stochastic Gradient Descent over Backbone Sizes}
\label{app:backbone-size}
\begin{figure}[tb]
    \centering
    \includegraphics[width=\linewidth]{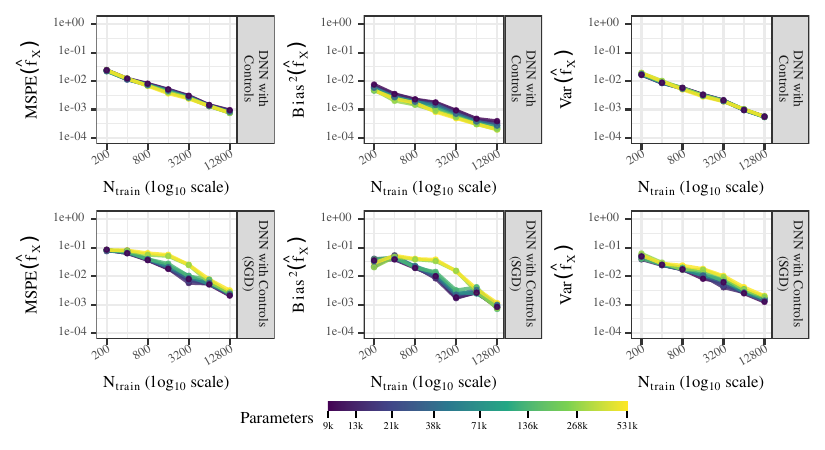}
    \caption{Convergence of $\hat f_X$ for the DNN with Controls as the backbone size grows, fit by refit (top, $2$-fold cross-fit) and end to end by SGD (bottom).
    Columns show $\mathrm{MSPE}$, $\mathrm{Bias}^2$, and $\mathrm{Var}$ of $\hat f_X$.
    Curves are coloured by the trainable parameter count of the DNN backbone, ranging from $\sim$9k to $\sim$531k.}
    \label{fig:concurvity_q}
\end{figure}

We compare two ways of fitting the DNN with Controls over increasing backbone size: our refit of the last layer, and fitting the full model end to end with stochastic gradient descent (SGD) (\cref{fig:concurvity_q}).
The refit is robust to backbone size: its curves collapse onto a single trajectory.
In contrast, SGD generally converges more slowly with increasing $N_\mathrm{train}$, and its $\mathrm{MSPE}$ remains higher than that of the refit at all sample sizes.
While SGD converges fairly fast for small backbones (darker lines), the relation between backbone size and convergence is more volatile, which might indicate a strong sensitivity to hyperparameter choice.
Hyperparameters are searched per method and along $(n,q)$ as documented in \cref{app:sim_extra}.
The size of the backbone was varied by adjusting the last-layer feature dimension $q$ from $8$ to $1024$, as done in \cref{sec:sim}.

\subsection{Noisy Control Variables}
\label{app:noisyz}

A control variable may be available only as a noisy measurement, for example when it is itself the output of a previous estimation step.
In the measurement error literature, the resulting phenomenon is \emph{attenuation}: the effect of the noisily observed variable is underestimated \cite{carroll2006MeasurementErrorNonlinear}.
In our setting, an attenuated covariate effect implies that the image effect is only partially corrected for omitted variable bias.
To quantify this, we repeat the simulation of \cref{sec:sim} with a continuous outcome and $\beta_Z = 1$, generating $X$ and $y$ from the true covariate $Z$, while all estimators receive only the corrupted version $\tilde Z = (1-a)\,Z + a\,\varepsilon$ with $\varepsilon \sim U([0,1])$ drawn independently of $Z$, for training, validation, and prediction.
The parameter $a \in \{0, 0.1, \ldots, 1\}$ interpolates between noise-free covariates ($a=0$) and covariates that are pure noise ($a=1$).
All other settings, including the training protocol and hyperparameters, are identical to the main simulation study (\cref{app:sim_extra}), with $n \in \{400, \ldots, 25{,}600\}$ and $50$ simulation replications per setting.

\begin{figure}[tb]
  \centering
  \subfloat[Squared bias.\label{fig:noisyz-bias}]{%
    \includegraphics[width=0.49\linewidth]{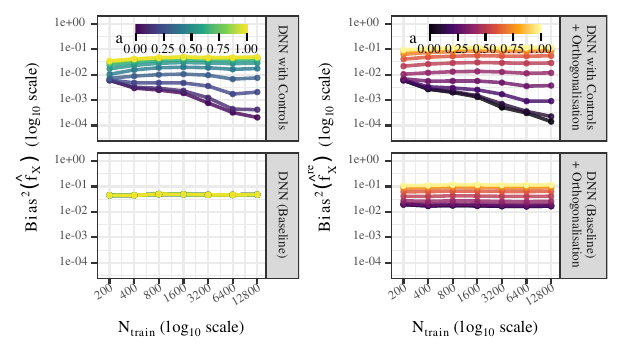}}
  \hfill
  \subfloat[Variance.\label{fig:noisyz-var}]{%
    \includegraphics[width=0.49\linewidth]{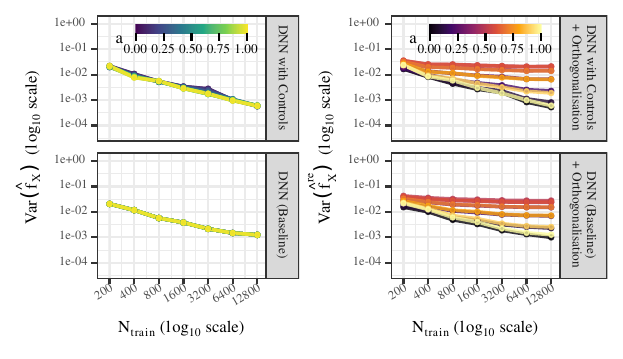}}
  \caption{Squared bias (a) and variance (b) of $\hat f_X$ (left blocks) and of the orthogonalised $\hat f_X^\mathrm{re}$ (right blocks) when estimators observe only the corrupted covariate $\tilde Z = (1-a)Z + a\varepsilon$, for noise levels $a \in \{0, 0.1, \ldots, 1\}$ (colour), over the sample size $N_\mathrm{train} = n/2$ available to the backbone and refit stages of each fold under $2$-fold cross-fitting.
  Top rows: DNN with Controls; bottom rows: DNN fit without covariates, with post-hoc orthogonalisation \cite{t.weber2023UnreadingRacePurging} for $\hat f_X^\mathrm{re}$.
  The bias interpolates between the noise-free decay at $a=0$ and the flat level of the uncontrolled DNN at $a=1$, while the variance of the DNN with Controls is essentially unaffected by $a$.}
  \label{fig:noisyz}
\end{figure}

\Cref{fig:noisyz-bias} shows the squared bias of $\hat f_X$.
For noise-free covariates, the DNN with Controls reproduces the behaviour observed in \cref{fig:bz}: the bias decays as the sample grows.
As $a$ increases, the curves flatten onto plateaus whose level rises with the noise, and at $a=1$ the plateau coincides with that of the uncontrolled DNN.
Controlling for a noisy covariate therefore removes part of the omitted variable bias and degrades gracefully to the uncontrolled model in the pure-noise limit, rather than beyond it.
The plateaus are flat in the sample size, so the remaining bias is not a small-sample artefact: more data does not compensate for a noisy control variable, in line with classical measurement error results \cite{carroll2006MeasurementErrorNonlinear}.
The same pattern holds for the orthogonalised estimator of the residual effect.
\Cref{fig:noisyz-var} shows the corresponding variances, which are essentially unaffected by $a$ for the DNN with Controls, so the noise level moves the bias--variance trade-off through the bias alone.
We discuss the practical implications of this limitation in \cref{sec:limitations}.

\subsection{Additional Details on the UK Biobank Experiment}
\label{app:ukbb-extra}

T1-weighted volumes were skull-stripped with using FreeSurfer \cite{FISCHL2012774} and rigidly registered to MNI152 space at $2\,$mm isotropic resolution.
 At training time, intensities are rescaled to $[0,1]$ within the brain mask; no further normalisation, augmentation, or masking is applied.

For confounding strength $\beta_\mathrm{age} \in \{0, 2\}$ and fixed $\beta_\mathrm{sex} = 2$ we draw synthetic covariates $age_i \sim \mathcal{N}(\mu_\mathrm{age}, 0.9\,\sigma_\mathrm{age})$ with $\mu_\mathrm{age}$ the sample mean and $\sigma_\mathrm{age}$ the sample variance over the training data.
The factor $0.9$ avoids over-sampling at the tails of the empirical age distribution.
We sample $sex$ uniformly with $sex_i \sim \mathrm{Bernoulli}(0.5)$.
We then sample the response $highalc_i$ from a Bernoulli with logit
$\beta_\mathrm{sex}\,(sex_i - 0.5) - z\_coef \cdot (age_i - \mu_\mathrm{age}) / (0.9\,\sigma_\mathrm{age})$.
Together, the three draws form a synthetic sample $(y, sex, age)$.
Then, each synthetic observation is matched to the closest real $(y, sex, age)$ row in the candidate fold by nearest-neighbour matching with replacement; a candidate may appear multiple times per fold, but the folds themselves are disjoint as resampling is applied per fold and after splitting the original training dataset into folds.
Per-fold training subsample size is $N = 5{,}000$ and test subsample size is $N_\mathrm{test} = 2{,}500$.

The 3D ResNet-50 backbone~\cite{hara2018CanSpatiotemporalD} ($2{,}048$ last-layer features) is fine-tuned per fold with binary cross-entropy and balanced class weights.
Hyperparameters were selected by random grid search over learning rate and weight decay.
Optimisation used Adam with learning rate $1.67 \times 10^{-6}$, weight decay $1.02 \times 10^{-5}$, batch size $48$, ReduceLROnPlateau scheduling (factor $0.8$, patience $5$), and early stopping on validation loss with patience $20$.
Model comparison uses five-fold stratified outer cross-validation; within each fold, the synthetic resampling and the train/refit partition for sample-split / cross-fit estimators are drawn independently, always ensuring that no samples leak across folds or train/refit partition inside each fold.

\begin{figure}[!htb]
    \centering
    \includegraphics[width=\linewidth]{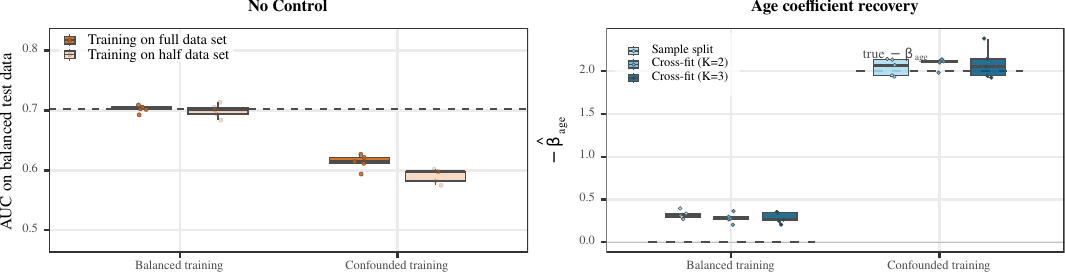}
    \caption{Additional diagnostic results for the UK Biobank application.
    \emph{(left)} Test-set AUC of the base DNN, full versus half training fold;
    \emph{(right)} coefficient recovery: estimated $-\hat \beta_\mathrm{age}$ vs.\ ground truth (dashed).
    All boxplots span five cross-validation folds.}
    \label{fig:ukbb-app}
\end{figure}

\Cref{fig:ukbb-app} reports additional results.
In general, we can see that $age$-marginalisation of a DNN with controls \emph{nearly} recovers the predictive performance of a DNN trained on balanced data, when using cross-fitting.
However, a small drop remains.
The first panel checks that reducing the backbone training set to half (the partition used by the post-hoc fits) does not by itself explain this gap.
Cross-fitting closes most of this residual sample-split AUC gap and reduces fold-to-fold variance.
The second panel reports the recovery of the synthetic covariate coefficient $\hat\beta_\mathrm{age}$ for the estimators using sample splitting.
While coefficients are nearly recovered, we can also observe a gap here.

The gaps are likely an artefact of the simulation study, as we use a real (not synthetic) target with $highalc$ — which could already carry some age- and sex-related signal of its own (see \cite{rane2022StructuralDifferencesAdolescent}.
The recovered coefficients could absorb both the injected confound and any pre-existing real-data covariate–outcome correlation.
A second factor is the non-collapsibility of logistic regression \cite{greenland1999ConfoundingCollapsibilityCausal} — the image effect plays the role of a brain-specific intercept, so the conditional coefficient identified after image control is generally not equal to the marginal data-level coefficient implied by the resampling, and the recovered estimate reflects this.
Finally, finite data means that the resampling procedure will oversample certain observations, especially in certain subpopulations of the dataset (e.g.\ female, high alcohol, older).
More cross-fitting folds naturally lead to smaller sample sizes per-fold, and as we apply resampling within each fold this leads to a stronger dataset bias per fold.

\subsection{Other Methods in the UK Biobank Experiment}
\label{app:ukbb-others}

All methods below are fit on the same training sample of $N = 5{,}000$ per fold as the DNN with Controls, with $Z = age$, and are evaluated on the same balanced test set.
\cref{tab:ukbb-others} reports their test AUC.
Where a method's predictor depends on $age$, we report its marginalised prediction (\cref{def:controlled-pred}) over the training $age$ distribution, $\hat y(x_i) = \frac{1}{N}\sum_{j=1}^{N} \sigma\big(\hat\eta(x_i, age_j)\big)$ to ensure a fair comparison.
PHO + Refit and LEACE both start from the last-layer features $\phi$ of DNN (Baseline).
We centre $\phi$ and $age$ on the training sample, giving $\phi_c$ and $z_c$, remove the $age$ information from $\phi_c$ as described below, and refit the last layer on the result.

\begin{table}[t]
\centering
\small
\caption{AUC on the age-balanced test set, mean (standard deviation) over 5 cross-validation folds. AUC (marg.) indicates AUC of the marginalised prediction (\cref{def:controlled-pred}). We denote the highest value per column by $^*$ and bold any results not significantly below the highest value (paired $t$-test over folds, $\alpha = 0.05$).}
\label{tab:ukbb-others}
\begin{tabular}{lcccc}
\toprule
 & \multicolumn{2}{c}{Balanced training} & \multicolumn{2}{c}{Confounded training} \\
Method & AUC & AUC (marg.) & AUC & AUC (marg.) \\
\midrule
DNN (Baseline) & 0.702 (0.006) & -- & 0.614 (0.013) & -- \\
DNN with Controls, no sample split & 0.690 (0.009) & 0.691 (0.009) & 0.611 (0.012) & 0.635 (0.014) \\
DNN with Controls, sample split & 0.678 (0.017) & 0.677 (0.017) & 0.569 (0.004) & \textbf{0.666 (0.018)} \\
DNN with Controls, cross-fit (K=2) & \textbf{0.704 (0.014)} & \textbf{0.701 (0.014)} & 0.569 (0.005) & \textbf{0.685 (0.012)} \\
DNN with Controls, cross-fit (K=3) & \textbf{0.714 (0.009)}$^*$ & \textbf{0.710 (0.009)}$^*$ & 0.560 (0.011) & \textbf{0.686 (0.003)}$^*$ \\
Joint DNN & \textbf{0.708 (0.013)} & \textbf{0.708 (0.013)} & 0.612 (0.011) & 0.612 (0.011) \\
DNN (Baseline) + Orth. & 0.703 (0.006) & \textbf{0.702 (0.006)} & 0.631 (0.016) & 0.614 (0.013) \\
PHO + Refit & 0.694 (0.007) & 0.694 (0.007) & 0.638 (0.015) & 0.624 (0.012) \\
LEACE & 0.693 (0.008) & 0.693 (0.008) & 0.642 (0.004) & 0.642 (0.004) \\
DFR & 0.645 (0.057) & 0.645 (0.057) & \textbf{0.672 (0.010)}$^*$ & 0.672 (0.010) \\
\bottomrule
\end{tabular}
\end{table}

\paragraph{Joint DNN.}
The joint DNN $g(X, Z)$ receives $age$ as an additional input: its $2{,}048$ backbone features are concatenated with standardised $age$ and passed through an MLP head with two hidden layers of width $64$ and ReLU activations.
The backbone starts from the same pre-trained initialisation as DNN (Baseline), and the model is trained end to end with balanced class weights and Adam (learning rate $10^{-5}$, weight decay $10^{-5}$).
The learning rate is halved by ReduceLROnPlateau after $3$ epochs without improvement, and training stops early on the validation loss of a stratified $80/20$ split (patience $15$, at most $100$ epochs).
Since $g$ is not additive, it has no component $f_X$ that could be separated from $f_Z$, and we therefore report the marginalised prediction $\frac{1}{N}\sum_j \sigma(g(x_i, age_j))$.

\paragraph{DNN (Baseline) + Orthogonalisation \cite{rugamer2023NewPHOrmulaImproved}.}
This method keeps the predictor $\hat\eta = \hat\beta_0 + \hat f_X(x)$ of DNN (Baseline) and orthogonalises it with respect to $age$ after training, that is, it removes its linear dependence on $age$ on the logit scale.
For this, we estimate $\hat b = \arg\min_b \lVert \hat f_X - z_c b \rVert^2$ on the training sample and use $\hat\eta(x, z) = \hat\beta_0 + \hat f_X(x) - z_c \hat b$.
Because the subtracted term depends on the $age$ of the observation, we report the marginalised prediction.

\paragraph{PHO + Refit \cite{t.weber2023UnreadingRacePurging}.}
Instead of the predictor, PHO + Refit residualises the last-layer features on $age$, $R = \phi_c - z_c \hat B$ with $\hat B = (z_c^\top z_c)^{-1} z_c^\top \phi_c$, and fits a new logistic last layer on $R$, as in \cite{t.weber2023UnreadingRacePurging}.
Unlike \cite{t.weber2023UnreadingRacePurging}, we add an $\ell_2$ penalty, since an unpenalised logistic regression refit on $2{,}048$ features and $5{,}000$ observations overfits and is unstable.
We standardise $R$ and choose the penalty strength $C^{-1}$ from $13$ log-spaced values $C \in [10^{-5}, 10]$ by the validation log-loss on a stratified $80/20$ split, and we then refit the layer on the full training sample.
At test time, we residualise the features with the training coefficients $\hat B$ and the $age$ of each observation, and we therefore report the marginalised prediction.

\paragraph{LEACE \cite{norabelrose2023LEACEPerfectLinear}.}
Least-squares concept erasure replaces $\phi_c$ by $r(\phi) = \phi_c - W^{+} P W \phi_c$, where $W = \Sigma_{\phi\phi}^{-1/2}$ whitens the features and $P$ is the orthogonal projection onto the column space of $W \Sigma_{\phi z}$.
Among all affine maps that make the features linearly uncorrelated with $age$, this map changes $\phi$ least in expected squared Euclidean distance.
The last layer is then fit on $r(\phi)$ in the same way as for PHO + Refit.
Since the erasure does not use $age$ at test time, no marginalisation is needed.

\paragraph{Deep feature reweighting (DFR) \cite{kirichenko2023last}.}
DFR retrains the last layer of a backbone on group-balanced subsets of the data, which keep all observations of the smallest group and subsample every other group to the same size.
Since DFR requires a discrete spurious attribute, we form $10$ groups from the combinations of the outcome $y$ and the $age$ quintile.
We follow the implementation of \cite{kirichenko2023last} with its default settings, in which the last layer is an $\ell_1$-penalised logistic regression on standardised features and the final weights are averaged over fits on independent balanced subsets.
DFR requires a training set for the backbone and a separate validation set for the last layer, so we use the backbone trained on one half of the training sample, as for Sample split, and the other half as validation set.
The penalty parameter $C \in \{1, 0.7, 0.3, 0.1, 0.07, 0.03, 0.01\}$ is chosen by worst-group accuracy on one random half of the validation set after fitting on a balanced subset of the other half, and the final layer averages $20$ fits on balanced subsets of the full validation set, with features standardised by the statistics of the training set.
Since DFR does not use $age$ at test time, no marginalisation is needed.

\subsection{Computational Resources}
\label{app:compute}

All experiments run on a compute server with $2\times28$ physical cores ($112$ hardware threads) at $2.0$\,GHz, $502$\,GB of system RAM, and two NVIDIA A100 80\,GB PCIe GPUs.
The simulation study runs on a single A100 with a pool of $4$ workers that processes all simulation settings, and takes approximately $28$\,h of active compute time in total.
In the UK Biobank application, a single outer fold (3D ResNet-50, $N_\mathrm{train}=5{,}000$) takes approximately $40$\,min for $K=2$ and $100$\,min for $K=3$, so that the five outer folds and two confounding settings require roughly $7$\,h of single-GPU compute for $K=2$ and $17$\,h for $K=3$.

\clearpage

\end{document}